\documentclass[12pt]{article}
\usepackage[utf8]{inputenc}
\usepackage{graphicx}
\usepackage{tabularx}
\usepackage{scalerel}
\usepackage{graphicx}
\usepackage{amsmath, amssymb}
\usepackage{blindtext}
\usepackage{subcaption}
\usepackage{caption}
\usepackage{tabu}
\usepackage{moresize}
\usepackage{apacite}
\usepackage{natbib}
\usepackage{setspace}
\usepackage{hyperref}
\usepackage{lscape}
\usepackage{colortbl,hhline}
\usepackage{hyperref}
\usepackage{times}
\usepackage{setspace}
\usepackage{chngcntr}
\usepackage{booktabs,multirow}
\usepackage{kbordermatrix}
\numberwithin{figure}{section}
\numberwithin{equation}{section}
\counterwithin{table}{section}
\usepackage{times}
\newenvironment{sciabstract}{%
	\begin{quote} \bf}
	{\end{quote}}
\title{Random cohort effects and age groups dependency structure for mortality modelling and forecasting: {Mixed-effects time-series model approach}} 

\author
{Ka Kin Lam\footnote{Corresponding author. E-mail: kakin.lam@abdn.ac.uk (K.K.Lam).},  Bo Wang\\
	\\
	\normalsize{School of Mathematics and Actuarial Science, University of Leicester,}\\
	\normalsize{Leicester LE1 7RH, UK}\\
}
\date{}

\begin{document} 
\baselineskip24pt
\maketitle 
\begin{sciabstract}
\begin{center}
Abstract
\end{center}
\normalfont There have been significant efforts devoted to solving the longevity risk given \textcolor{black}{that} a continuous growth in population ageing has become a severe issue for many developed countries over the past few decades. The Cairns-Blake-Dowd (CBD) model, which incorporates cohort effects parameters in its parsimonious design, is one of the most well-known approaches for mortality modelling \textcolor{black}{at higher ages and longevity risk}. This article \textcolor{black}{proposes a novel mixed-effects time-series approach for mortality modelling and forecasting with considerations of age groups dependence and random cohort effects parameters.} The proposed model can disclose more mortality data information and provide a natural quantification of the model parameters uncertainties with no pre-specified constraint required for estimating the cohort effects parameters. The abilities of the proposed approach are demonstrated through two applications with empirical male and female mortality data. The proposed approach shows remarkable improvements in terms of forecast accuracy compared to \textcolor{black}{the CBD model} in the short-, mid-and long-term forecasting using mortality data of several developed countries in the numerical examples.
\end{sciabstract}
\section{Introduction}
Continuous growth in life expectancy has been lasting over the last few decades in many developed countries. This unanticipated increase in lifespan has posed significant challenges to many government sectors and life insurance sectors due to the extra burdens on the health care services and the rapid increase in pension and insurance expenditure. The problem caused by the ageing population is also well-known as the `longevity risk'. Many demographic researchers have recognised this problem and devoted significant efforts to developing new statistical models for mortality modelling and forecasting. A good statistical model for modelling mortality is always of great importance to incorporate many different factors that help explain the historical mortality patterns and capture their future trends. For example, \textcolor{black}{cohort effects} in a statistical mortality model can be considered one of the most important factors but is still underexplored and is missing from the majority of mortality modelling literature, which we aim to discuss and address in this article.
\par
Cohort effects, also known as the year-of-birth effects, are arguably one of the most important and discussed factors that influence mortality experience among groups of individuals born in the same year or generation. Some evidence for the existence of the cohort effects has been successfully spotted, and the developments of the cohort studies can mainly be divided into two different streams. One side relies on the empirical data analysis and the qualitative analysis, such as descriptive and graphical representations, to conclude the existence of cohort effects on mortality patterns in a population. For example, \cite{willets2004cohort} discovers \textcolor{black}{that} the `golden generations' cohort who were born between the year 1925 and the year 1945, have experienced exceptionally rapid improvements in mortality rates through examining the average annual mortality improvement rates for males and females in the population of England and Wales. \cite{murphy2009golden} and \cite{murphy2010reexamining} further explain this `golden generations' cohort phenomenon that may be due to those born during and after the Second World War with low fertility and thus facing less competition for resources, better environmental conditions with changing smoking habits and diets between generations. \cite{li2016two} demonstrate the cohort surface across five different countries using graphical presentations with kernel smoothing techniques and conclude that Japan and the UK are the countries with relatively stronger cohort effects than the Netherlands, Spain and the US.
\par Another side of the cohort studies depends on establishing a new statistical model that offers quantitative approaches to identify the cohort patterns of the observed mortality data. Early attempts can be traced back to the classical age-period-cohort (APC) model in mortality modelling \citep{holford1983estimation}. \cite{renshaw2003lee} extend the well-known Lee-Carter model \citep{lee1992modeling} by incorporating an age-modulated cohort effects parameter. \cite{cairns2009quantitative} extend their original two-factor Cairns-Blake-Dowd (CBD) model \citep{cairns2006two} as a three-factor CBD model with cohort parameters. However, the aforementioned statistical mortality models suffer the `age-period-cohort identification problem', given that the age, period and cohort effects are intertwined\footnote{The age $x$ and period $t$ take values in $\{x_{1},\cdots, x_{m}\}$ and $\{ t_{1},\cdots, t_{n}\}$, the cohort index $t-x$ takes values in $\{t_{1}-x_{m},\cdots,t_{n}-x_{1}\}$.}. Including indicator variables of each level of age, period and cohort respectively in a fixed-effects estimation approach can lead to the perfect collinearity problem, which means that there can be an infinite number of solutions that fit the observed data equally well in the fixed-effects estimation approach \citep{o2014estimable}. The common solution to this age-period-cohort identification problem relies on placing pre-specified constraints on the model coefficients prior to the fitting procedure. However,  such pre-specified constraints meanwhile lead to the fitting results of the estimated model parameters substantially depending on the initial constraints chosen \citep{o2017mixed}. Readers can refer \textcolor{black}{to} \cite{haberman2011comparative} and \cite{currie2016fitting} for more comprehensive summaries on these convergence and robustness problems among all these fixed-effects estimation approaches applied to the morality models with the cohort effects parameters. 
\par In this article, we focus on the side of establishing a new statistical mortality model that can extract and provide cohort effects from the observed mortality data for modelling and forecasting. We propose \textcolor{black}{a novel approach} with random cohort effects parameters and age groups dependency structure, which can be seen as a mixed-effects time-series version of the three-factor CBD model with the cohort effects parameters proposed by \cite{cairns2009quantitative}. The main contributions of this article are to demonstrate the formulation and parameters estimation of \textcolor{black}{the proposed model} with considerations of random cohort effects parameters for mortality modelling and forecasting under a united mixed-effects time-series framework with several desirable advantages over \textcolor{black}{the CBD model}. 
\par More will be discussed in detail and the rest of this article is organised as follows. In Section \ref{Section 2: time-series variant mixed-effects CBD model}, we firstly give a review of \textcolor{black}{the CBD model}. In Section \ref{Section 3: time-series variant mixed-effects CBD model}, we introduce the formulation and the framework of \textcolor{black}{the proposed model} for mortality modelling and forecasting. We then illustrate the proposed model with empirical studies of mortality modelling and forecasting with comparisons to \textcolor{black}{the CBD model} in terms of the systematic differences and forecasting ability using mortality data of ten different developed countries in Section \ref{Section 4: time-series variant mixed-effects CBD model}. We lastly conclude this article with discussions and remarks in Section \ref{Section 5: time-series variant mixed-effects CBD model }.

\section{Review of the Cairns-Blake-Dowd (CBD) model} \label{Section 2: time-series variant mixed-effects CBD model}
In this section, we define some common actuarial notations \textcolor{black}{used} in the CBD model first, \textcolor{black}{and} then go through a brief review of the CBD model. 
\subsection{Actuarial notations and their relationships}
In the literature of mortality modelling, there are generally two types of stochastic methods. One type of method models the central mortality rate. The CBD model belongs to \textcolor{black}{the other} type of method which models the initial mortality rate. For the sake of notation clarity, we start by defining the notations using a subscript $i$ to denote the number of calendar year $t_{i}$ for $i = 1,\cdots,n$, and a subscript $j$ as the number of age-specific group $x_{j}$ for $j = 1,\cdots,m$ in the CBD model. These notations include as follows:
\begin{itemize}
	\item {$D_{x_{j},t_{i}}$ is the observed number of deaths aged $x_{j}$ last birthday in a calendar year $t_{i}$.}
	\item {$E^{c}_{x_{j},t_{i}}$ is the central exposure to risk aged $x_{j}$ last birthday in the middle of a calendar year $t_{i}$.}
	\item {$m_{x_{j},t_{i}}$ is the central mortality rate which reflects the death
	probability within the range of 0 to 1 for a population died in age $x_{j}$ last birthday in the middle of a calender year $t_{i}$. It is estimated by $m_{x_{j},t_{i}} = D_{x_{j},t_{i}}/E^{c}_{x_{j},t_{i}}$.}
	\item {$q_{x_{j},t_{i}}$ is the initial mortality rate as the one-year death probability between $0$ and $1$ for a population died in age $x_{j}$ exactly at time $t_{i}$.}
\end{itemize}
The initial mortality rate $q_{x_{j},t_{i}}$ and the central mortality rate $m_{x_{j},t_{i}}$ are typically very close to one another in practice \citep{dickson2013actuarial}. The link between the initial mortality rate and the central mortality rate can be stated, i.e.
\begin{equation}\label{eq: the relationship among q and m}
	q_{x_{j},t_{i}} \thickapprox 1 - \text{exp}(-m_{x_{j},t_{i}}),
\end{equation}
where $0 \leq q_{x_{j},t_{i}} \leq 1$.

\subsection{Cairns-Blake-Dowd (CBD) model}
The Cairns-Blake-Dowd (CBD) model is one of the most well-known approaches in the mortality modelling literature. \cite{cairns2009quantitative} extend their original two-factor CBD model and propose a three-factor CBD model which incorporates the cohort parameters based on the parsimonious design in their previous version. 
\par We begin by using a subscript $i$ to denote the number of a calendar year $t_{i}$ for $i = 1,\cdots,n$, and a subscript $j$ for the number of an age-specific group $x_{j}$ for $j = 1,\cdots,m$\textcolor{black}{. The} CBD model has the form
\begin{equation}\label{eq: CBD model}
	y_{x_{j},t_{i}} = \text{logit}(q_{x_{j},t_{i}}) =  \text{log}\bigg(\frac{q_{x_{j},t_{i}}}{1 - q_{x_{j},t_{i}}}\bigg) = \kappa^{(1)}_{t_{i}} + \textcolor{black}{\kappa^{(2)}_{t_{i}}\times(x_{j} - \bar{x})} + \gamma^{(3)}_{t_{i}-x_{j}},
\end{equation}
where $y_{x_{j},t_{i}}$ is the logit of the initial mortality rate at age $x_{j}$ in a calendar year $t_{i}$, $\kappa_{t_{i}}^{(1)}$ is the general time effects intercept parameter, $\kappa_{t_{i}}^{(2)}$ is the time effects slope parameter, $\bar{x}$  is the average of the observed mortality data age range, i.e. $\bar{x} = \frac{1}{m}\sum_{j=1}^{m}x_{j}$, and $\gamma^{(3)}_{t_{i}-x_{j}}$ is the cohort effects parameter that provides mortality information on a group of people born in the same calendar year.
\par The CBD model relies on the linearity of the logit of the initial mortality rate at older ages (age 60 to age 89) by assuming that $y_{x_{j},t_{i}}$ is a response variable of a linear function. It treats $\kappa^{(1)}_{t_{i}}$ as an
intercept term, $\kappa^{(2)}_{t_{i}}$ as a slope term, and $\gamma^{(3)}_{t_{i}-x_{j}}$ as a cohort effects term and assumes all of the terms following stochastic processes for any given calendar year $t_{i}$ for mortality modelling \citep{cairns2009quantitative}.

\subsubsection {Model parameters estimation via the generalised linear modelling approach}
Instead of assuming the model error structure is normally distributed and homoscedastic as in many ordinary least squares linear regression approaches, the parameters in the CBD model are estimated based on an alternative approach known as the generalised linear modelling (GLM) approach to calibrate the historical mortality data to the model parameters. 
\par We start with the explanation of the CBD model parameters estimation procedure using the relationship between the initial mortality rate $q_{x_{j},t_{i}}$ and the central mortality rate $m_{x_{j},t_{i}}$ defined in Equation (\ref{eq: the relationship among q and m}), and the representation of the CBD model in Equation (\ref{eq: CBD model}). It can be denoted as  
\begin{equation}\label{eq: CBD model 2}
	\text{log}\bigg(\frac{1 - \text{exp}(-m_{x_{j},t_{i}})}{1 - (1 - \text{exp}(-m_{x_{j},t_{i}}))}\bigg) = \kappa^{(1)}_{t_{i}} + \textcolor{black}{\kappa^{(2)}_{t_{i}}\times(x_{j} - \bar{x})} + \gamma^{(3)}_{t_{i}-x_{j}}.
\end{equation}
Rearranging the order of Equation (\ref{eq: CBD model 2}) with respect to $m_{x_{j},t_{i}}$, we have
\[
	m_{x_{j},t_{i}} = \text{log}(1 + \text{exp}\textcolor{black}{(\kappa^{(1)}_{t_{i}} + \kappa^{(2)}_{t_{i}}\times(x_{j} - \bar{x}) + \gamma^{(3)}_{t_{i}-x_{j}})}).
\]
We denote above $m_{x_{j},t_{i}}$ as $m_{x_{j},t_{i}}(\theta_{i})$, where  $\theta_{i} = (\kappa_{t_{i}}^{(1)},\kappa_{t_{i}}^{(2)},\textcolor{black}{\gamma^{(3)}_{t_{i}-x_{j}}})$ represents the set of the model parameters depending on $m_{x_{j},t_{i}}$. The CBD model assumes that the number of deaths $D_{x_{j},t_{i}}$ is a random  variable from count data following the Poisson distribution, i.e.
\[
D_{x_{j},t_{i}} \sim \text{Poisson}(E^{c}_{x_{j},t_{i}} \times m_{x_{j},t_{i}}(\theta_{i})).
\]
We can thus state the log-likelihood function of $\boldsymbol{\theta}$  conditional on \textcolor{black}{$\boldsymbol{D}$ and $\boldsymbol{E^{c}}$} across all age-specific groups and calender years as
\begin{equation} \label{ch2eq: Orginal CBD model likelihood function}
	\mathcal{L}(\boldsymbol{\theta} \vert \textcolor{black}{\boldsymbol{D}, \boldsymbol{E^{c}}}) = \sum_{i=1}^{n}\sum_{j=1}^{m} D_{x_{j},t_{i}}\times\text{log}\textcolor{black}{(}E^{c}_{x_{j},t_{i}} \times m_{x_{j},t_{i}}(\theta_{i})\textcolor{black}{)} - E^{c}_{x_{j},t_{i}} \times m_{x_{j},t_{i}}(\theta_{i}) - \text{log}(D_{x_{j},t_{i}}!),
\end{equation}
\textcolor{black}{where $\boldsymbol{\theta}$ is the full set of the model parameters matrix, $\boldsymbol{D}$ is the observed number of deaths matrix across all age-specific groups and calender years, and $\boldsymbol{E^{c}}$ is the central exposure matrix across all age-specific groups and calender years.}

However, the full set of the model parameters cannot simply be calibrated by maximising the above log-likelihood function to have a closed form solution as the log-likelihood function follows the Poisson distribution. The model itself also suffers model identification problems, in which the model parameters are over-parametrised in Equation (\ref{eq: CBD model}) since there can be two or more possible model structures. For example, we can see that they are invariant to the following transformation
\[
(\kappa^{(1)}_{t_{i}}, \kappa^{(2)}_{t_{i}}, \gamma^{(3)}_{t_{i}-x_{j}}) \mapsto \Big(\kappa^{(1)}_{t_{i}}+\phi_{1}+\textcolor{black}{\phi_{2}\times(t_{i}-\bar{x})}, \kappa^{(2)}_{t_{i}} - \phi_{2}, \gamma^{(3)}_{t_{i}-x_{j}} - \phi_{1} - \textcolor{black}{\phi_{2}\times(t_{i} - x_{j})}\Big),
\]
where $\phi_{1}$ and $\phi_{2}$ are any non-zero constants.
\par \cite{cairns2009quantitative} amend this identification problem by switching the cohort effects parameter from $\gamma^{(3)}_{t_{i}-x_{j}}$ to  $\tilde{\gamma}^{(3)}_{t_{i}-x_{j}} = \phi_{1}+ \textcolor{black}{\phi_{2}\times(t_{i}-x_{j})} + \gamma^{(3)}_{t_{i}-x_{j}}$ and imposing the following constraints on the cohort parameter $\gamma^{(3)}_{t_{i}-x_{j}}$ throughout the iterative optimisation process for the log-likelihood function in Equation (\ref{ch2eq: Orginal CBD model likelihood function}), such that
\[ \sum_{(t_{i}-x_{j}) \in C} \gamma^{(3)}_{t_{i}-x_{j}} = 0, \]
\[ \sum_{(t_{i}-x_{j} ) \in C} (t_{i}-x_{j}) \times \gamma^{(3)}_{t_{i}-x_{j}} = 0, \]
where $C$ is the set of cohort years of birth $(t_{i}-x_{j})$ included in the analysis. \cite{cairns2009quantitative} explain the choice of the above constraints is analogous to fitting a latent linear function of $\phi_{1}+ \textcolor{black}{\phi_{2}\times(t_{i}-x_{j})}$ to $\gamma^{(3)}_{t_{i}-x_{j}}$ implicitly, so that the constraints ensure $\hat{\phi}_{1} = 0$ and $\hat{\phi}_{2} = 0$, and the fitted $\hat{\gamma}^{(3)}_{t_{i}-x_{j}}$ will fluctuate around zero and has no discernible linear \textcolor{black}{trend} as a standard normal random variable.  \cite{cairns2009quantitative}, \cite{haberman2011comparative} and \cite{currie2016fitting} provide further discussions on these identifiability and parameter fitting issues for the three-factor CBD model.
\subsubsection{Out-of-sample forecasts of the mortality odds in the CBD model}
For the out-of-sample forecasts of the initial mortality rate, \cite{cairns2006two} assume that the fitted $\hat{\kappa}_{t_{i}}^{(1)}$ and $\hat{\kappa}_{t_{i}}^{(2)}$ follow a bivariate random walk with drift which can be described by defining $\hat{\boldsymbol{\beta}}_{t_{i}} = \begin{pmatrix}\hat{\kappa}_{t_{i}}^{(1)} \\ \hat{\kappa}_{t_{i}}^{(2)}\end{pmatrix}$, then we have 
\[
\hat{\boldsymbol{\beta}}_{t_{i}} = \hat{\boldsymbol{\beta}}_{t_{i-1}} + \boldsymbol{d} + \boldsymbol{K}\boldsymbol{Z}_{\textcolor{black}{t_{i}}},
\]
where ${\boldsymbol{d}}$ is a $2 \times 1$ vector of the drift factors and $\boldsymbol{Z}_{\textcolor{black}{t_{i}}}$ is a $2 \times 1$ vector of bivariate normally distributed random variables, i.e. $\boldsymbol{Z}_{\textcolor{black}{t_{i}}} \sim \mathcal{N}_{2 \times 1}(\boldsymbol{0},\boldsymbol{I})$. Let $\boldsymbol{V}$ be a variance-covariance matrix of $\hat{\kappa}_{t_{i}}^{(1)}$ and $\hat{\kappa}_{t_{i}}^{(2)}$. Without loss of generality, $\boldsymbol{K}$ is restricted to be an upper triangular matrix by Cholesky decomposition where $\boldsymbol{V} = \boldsymbol{K}^{T}\boldsymbol{K}$. Denoting $\Delta$ as a first-difference operator, such that  $\Delta\hat{\kappa}_{t_{i}}^{(l)} =  \hat{\kappa}_{t_{i}}^{(l)} - \hat{\kappa}_{t_{i-1}}^{(l)}$ for $l = 1, 2$, we can estimate $\boldsymbol{d}$ and $\boldsymbol{V}$ as 
\[
	\boldsymbol{\hat{d}} =  \begin{pmatrix}  \mathbb{E}(\Delta\hat{\kappa}_{t_{i}}^{(1)}) \\ \mathbb{E}(\Delta\hat{\kappa}_{t_{i}}^{(2)}) \end{pmatrix},
\]
and let $\hat{\sigma}_{lk} = \mathbb{E}[(\Delta\hat{\kappa}_{t_{i}}^{(l)} -  \mathbb{E}(\Delta\hat{\kappa}_{t_{i}}^{(l)}))^{T}(\Delta\hat{\kappa}_{t_{i}}^{(k)} -  \mathbb{E}(\Delta\hat{\kappa}_{t_{i}}^{(k)}))]$ for $l,k = 1,2$,

\[
	\boldsymbol{\hat{V}} = \begin{pmatrix} \hat{\sigma}_{11} & \hat{\sigma}_{12} \\ \hat{\sigma}_{21} & \hat{\sigma}_{22} \end{pmatrix}.
\]
By Cholesky decomposition of $\hat{\boldsymbol{V}}$, we have
\[
	\boldsymbol{\hat{V}} = \boldsymbol{\hat{K}}^{T}\boldsymbol{\hat{K}}.
\]
The $h$-step ahead extrapolation of $\hat{\boldsymbol{\beta}}_{t_{i}}$ is then
\[
\textcolor{black}{	\hat{\boldsymbol{\beta}}_{t_{i+h}} = \hat{\boldsymbol{\beta}}_{t_{i+h-1}} + \boldsymbol{\hat{d}},}
\]
and the variance of the $h$-step ahead extrapolation of $\hat{\boldsymbol{\beta}}_{t_{i}}$ is
\[
{\text{Var}({\boldsymbol{\beta}}_{t_{i+h}}) = h \times \boldsymbol{\hat{V}}.}
\]

\par The cohort effects parameter $\hat{\gamma}^{(3)}_{\textcolor{black}{t_{i}-x_{j}}}$ is assumed to follow a univariate random walk with drift \citep{haberman2011comparative}, hence the $h$-step ahead extrapolation of $\hat{\gamma}_{\textcolor{black}{t_{i}-x_{j}}}^{(3)}$ is 
\[
\textcolor{black}{	\hat{\gamma}^{(3)}_{t_{i+h}-x_{j}} =  \hat{\gamma}^{(3)}_{t_{i+h-1}-x_{j}} + \hat{\mu} ,}
\]
where $\hat{\mu} = \mathbb{E}(\Delta\hat{\gamma}^{(3)}_{t_{i}-x_{j}})$.
\par Combining all the extrapolated model parameters in the CBD model, the $h$-step ahead out-of-sample forecast of $y_{x_{j},t_{i}}$ is
\[
	\hat{y}_{x_{j},t_{i+h}} = \hat{\kappa}^{(1)}_{t_{i+h}} + \textcolor{black}{\hat{\kappa}^{(2)}_{t_{i+h}} \times (x_{j} - \bar{x})} +  \hat{\gamma}^{(3)}_{t_{i+h}-x_{j}}.
\]
\section{Methodology} \label{Section 3: time-series variant mixed-effects CBD model}
\subsection{\textcolor{black}{Mixed-effects time-series model} with age groups dependency and random cohort effects parameters}
In this section, we introduce the proposed \textcolor{black}{mixed-effects time-series model} with age groups dependency and random cohort effects parameters.
\par 
Let our observed mortality dataset be $\{t_{i},x_{j}, y_{x_{j},t_{i}}\}$ for $i = 1,\ldots,n$ and $j = 1,\ldots,m$, where $n$ is the total number of observed calendar years, $m$ is the maximum number of age-specific groups of interest, $y_{x_{j},t_{i}}$ is the logit of the initial mortality rate for age $x_{j}$ in a calendar year $t_{i}$. The proposed \textcolor{black}{mixed-effects time-series model} is 
\[
y_{x_{j},t_{i}} = \text{logit}(q_{x_{j},t_{i}}) =\text{log}\bigg(\frac{q_{x_{j},t_{i}}}{1 - q_{x_{j},t_{i}}}\bigg) = f_{x_{j}}(t_{i} - \bar{t}) + \epsilon_{x_{j},t_{i}}, \hspace{0.5cm} \epsilon_{x_{j},t_{i}} \sim \mathcal{N}(0, \sigma^{2}),
\]
where $f_{x_{j}}(t_{i} - \bar{t})$ is an underlying function which needs to be estimated for a given age $x_{j}$, $\bar{t}$ is the average of the observed calendar years, i.e. $\bar{t} = \frac{1}{n}\sum_{i=1}^{n} t_{i}$, and $\{\epsilon_{x_{j},t_{i}}\}_{i,j=1}^{n,m}$ allow the observation error varying among the different age groups and the calendar years based on the assumption of i.i.d. normally distributed random variables with mean 0 and constant variance  $\sigma^{2}$.
\par The underlying function $f_{x_{j}}(t_{i} - \bar{t})$ in the proposed model under the mixed-effects structure \textcolor{black}{is further specified as}
\[
f_{x_{j}}(t_{i} - \bar{t}) = \mu(t_{i} - \bar{t}) + \eta_{x_{j}}(t_{i} - \bar{t}),
\] 
where $\mu(t_{i} - \bar{t})$ is a general linear mean function across all age groups $\{x_{j}\}_{j=1}^{m}$, i.e.
\[
\mu(t_{i} - \bar{t}) = \beta^{(1)} + \beta^{(2)}\times(t_{i} - \bar{t}),
\]
with $\beta^{(1)}$ is the fixed-effects intercept parameter and $\beta^{(2)}$ is the fixed-effects slope parameter multiplying the input $(t_{i} - \bar{t})$. \par Meanwhile, $\eta_{x_{j}}(t_{i} - \bar{t})$ is the age-specific deviation function from the general linear mean function across all age groups, which is designed to be a function with the following components
\begin{equation}\label{eq: age group specific deviation trend function}
\eta_{x_{j}}(t_{i} - \bar{t}) = f_{x_{j}}(t_{i} - \bar{t}) - \mu(t_{i} - \bar{t}) =  \gamma_{x_{j}}^{(1)} + \gamma_{x_{j}}^{(2)} \times (t_{i} - \bar{t}) + \gamma^{(3)}_{t_{i}-x_{j}},
\end{equation}
where we assume that \textcolor{black}{$\{\gamma_{x_{j}}^{(1)}\}_{j=1}^{m}$} are the correlated age-specific random effects intercept \textcolor{black}{parameters} following a prior distribution \textcolor{black}{${(\gamma_{x_{1}}^{(1)},\ldots,\gamma_{x_{m}}^{(1)})}^{T} \sim \mathcal{N}(\boldsymbol{0},\boldsymbol{K}^{(1)})$}, \textcolor{black}{$\{\gamma_{x_{j}}^{(2)}\}_{j=1}^{m}$} are the correlated age-specific random effects slope \textcolor{black}{parameters} following a prior distribution \textcolor{black}{${(\gamma_{x_{1}}^{(2)},\ldots,\gamma_{x_{m}}^{(2)})}^{T} \sim \mathcal{N}(\boldsymbol{0},\boldsymbol{K}^{(2)})$}, and \textcolor{black}{$\{\gamma_{t_{i}-x_{j}}^{(3)}\}_{t_{1} - x_{m}}^{t_{n} - x_{1}}$} are the correlated age-specific random effects cohort \textcolor{black}{parameters} following a prior distribution \textcolor{black}{$(\gamma_{t_{1} - x_{m}}^{(3)}, \ldots, \gamma_{t_{n} - x_{1}}^{(3)})^{T} \sim \mathcal{N}(\boldsymbol{0},\boldsymbol{K}^{(3)})$} \footnote{Without any loss of completeness, the covariance structures of all the random effects components $\gamma_{x_{j}}^{(1)}$,  ${\gamma_{x_{j}}^{(2)}}$ and $\gamma^{(3)}_{t_{i}-x_{j}}$ will be discussed in Section \ref{Estimation of the time-series variant mixed-effects CBD model with age groups dependency and random cohort effects structure}.}. 
\par With all the designed components mentioned above, we can state the full form of \textcolor{black}{the proposed model} as
\begin{equation}\label{eq: full form of the time variant CBD model}
	y_{x_{j},t_{i}} = \beta^{(1)} + \textcolor{black}{\beta^{(2)}\times(t_{i} - \bar{t})} + \gamma_{x_{j}}^{(1)} + \textcolor{black}{\gamma_{x_{j}}^{(2)}\times(t_{i} - \bar{t})} + \gamma^{(3)}_{t_{i}-x_{j}} + \epsilon_{x_{j},t_{i}}.
\end{equation}
Alternatively, we can group $\kappa_{x_{j}}^{(1)} = \textcolor{black}{\beta^{(1)} + \gamma_{x_{j}}^{(1)}}$ and $\kappa_{x_{j}}^{(2)} = \textcolor{black}{\beta^{(2)} + \gamma_{x_{j}}^{(2)}}$ for the simplicity of notation, \textcolor{black}{and the proposed model} is thus
\[
y_{x_{j},t_{i}} = \kappa_{x_{j}}^{(1)} + \textcolor{black}{\kappa_{x_{j}}^{(2)}\times(t_{i} - \bar{t})} + \gamma^{(3)}_{t_{i}-x_{j}} + \epsilon_{x_{j},t_{i}}.
\]

\subsection{Formulation and parameters estimation of \textcolor{black}{the proposed  model}} \label{Estimation of the time-series variant mixed-effects CBD model with age groups dependency and random cohort effects structure}
In this section, we begin by describing the formulation of \textcolor{black}{the proposed model} and explain the procedure of the model parameters estimation.
\subsubsection{Model formulation }
\par \textcolor{black}{Letting} $\boldsymbol{Y}_{{x}_{j}} = \textcolor{black}{[{y}_{x_{j},t_{1}},{y}_{x_{j},t_{2}},\ldots,{y}_{x_{j},t_{n}}]^{T}}$, $\boldsymbol{T}_{{x}_{j}} = [(t_{1} - \bar{t}),(t_{2} - \bar{t}),\ldots,(t_{n} - \bar{t})]^{T}$ and $\boldsymbol{\varepsilon}_{x_{j}} = \textcolor{black}{[\varepsilon_{x_{j},t_{1}},\varepsilon_{x_{j}, t_{2}},\ldots,\varepsilon_{x_{j},t_{n}}]^{T}}$, we can denote Equation (\ref{eq: full form of the time variant CBD model}) in matrix form as
\begin{equation}\label{eq: (matrix) age group specific deviation trend function}
	\boldsymbol{Y} = \boldsymbol{T}\boldsymbol{\beta} + \boldsymbol{Z}^{(1)}\boldsymbol{\gamma}^{(1)}_{\boldsymbol{x}} +  \boldsymbol{Z}^{(2)}\boldsymbol{\gamma}^{(2)}_{\boldsymbol{x}} +  \boldsymbol{Z}^{(3)}\boldsymbol{\gamma}^{(3)}_{\boldsymbol{t-x}} + \boldsymbol{\epsilon}, \hspace{0.5cm} \boldsymbol{\epsilon} \sim N(\boldsymbol{0},\sigma^{2}\boldsymbol{I}),
\end{equation}
where 
\[
\boldsymbol{Y}=
\begin{bmatrix} 
\boldsymbol{Y}_{{x}_{1}}\\ \boldsymbol{Y}_{{x}_{2}} \\ \vdots \\ \boldsymbol{Y}_{{x}_{m}}
\end{bmatrix}
\in \mathbb{R}^{nm\times 1}
,
\boldsymbol{T}=
\begin{bmatrix} 
\boldsymbol{1}_{n \times 1} & \boldsymbol{T}_{{x}_{1}}\\ \boldsymbol{1}_{n \times 1} & \boldsymbol{T}_{{x}_{2}}\\ \vdots &  \vdots \\ \boldsymbol{1}_{n \times 1} & \boldsymbol{T}_{{x}_{m}}
\end{bmatrix}
\in \mathbb{R}^{nm\times 2},
\boldsymbol{\beta}=
\begin{bmatrix} 
\beta^{(1)} \\ \beta^{(2)}
\end{bmatrix}
\in \mathbb{R}^{2\times 1},
\]
\[
\boldsymbol{Z}^{(1)} = 
\begin{bmatrix} 
\boldsymbol{1}_{n \times 1} & 0 & \hdots & 0 & 0 \\
0&  \boldsymbol{1}_{n \times 1} & \hdots & 0 & 0 \\
\vdots &  \vdots &   \ddots &  \vdots & \vdots \\
0 & 0 &  \hdots &  0 &\boldsymbol{1}_{n \times 1} \\
\end{bmatrix}
\in \mathbb{R}^{nm\times m}
,
\boldsymbol{\gamma}^{(1)}_{\boldsymbol{x}} = 
\begin{bmatrix}
\gamma^{(1)}_{x_{1}}\\
\gamma^{(1)}_{x_{2}}\\
\vdots\\
\gamma^{(1)}_{x_{m}}\\
\end{bmatrix}
\in \mathbb{R}^{m\times 1},
\] 
\\
\[
\boldsymbol{Z}^{(2)} = 
\begin{bmatrix} 
\boldsymbol{T}_{x_{1}} & 0 & \hdots & 0 & 0 \\
0&  \boldsymbol{T}_{x_{2}} & \hdots & 0 & 0 \\
\vdots &  \vdots &   \ddots &  \vdots & \vdots \\
0 & 0 &  \hdots &  0 &\boldsymbol{T}_{x_{m}} \\
\end{bmatrix}
\in \mathbb{R}^{nm\times m},
\boldsymbol{\gamma}^{(2)}_{\boldsymbol{x}} = 
\begin{bmatrix}
\gamma^{(2)}_{x_{1}}\\
\gamma^{(2)}_{x_{2}}\\
\vdots\\
\gamma^{(2)}_{x_{m}}\\
\end{bmatrix}
\in \mathbb{R}^{m\times 1},
\] 
\\
\[
\boldsymbol{Z}^{(3)} = \kbordermatrix{
	& t_{1} - x_{m} & t_{2} - x_{m} & \hdots& t_{n} - x_{m}& t_{n} - x_{m-1} & \hdots & t_{n} - x_{1} \\
	t_{1} - x_{1} & 0 & 0 &  \hdots & 0  & 0 & \hdots & 0\\
	t_{2} - x_{1}  & 0 & 0 &  \hdots & 0 & 0 & \hdots & 0\\
	\vdots & \vdots & \vdots & \ddots & \vdots  & \vdots & \ddots  & \vdots \\
	t_{n} - x_{1}& 0 & 0 &   \hdots & 0 & 0 & \hdots & 1 \\
	t_{1} - x_{2} & 0 & 0 &  \hdots & 0 & 0 & \hdots & 0 \\
	t_{2} - x_{2} & 0 & 0 &  \hdots & 0 & 0 & \hdots & 0 \\
	\vdots & \vdots & \vdots & \ddots & \vdots  & \vdots & \ddots  & \vdots \\
	t_{n} - x_{2} & 0 & 0 &  \hdots & 0 & 0 & \hdots & 0 \\
	\vdots & \vdots & \vdots & \ddots & \vdots  & \vdots & \ddots  & \vdots \\
	t_{1} - x_{m-1} & 0 & 0 &  \hdots & 0 & 0 & \hdots & 0 \\
	t_{2} - x_{m-1} & 0 & 0 &  \hdots & 0 & 0 & \hdots & 0 \\
	\vdots & \vdots & \vdots & \ddots & \vdots  & \vdots & \ddots  & \vdots \\
	t_{n} - x_{m-1} & 0 & 0 &  \hdots & 0 & 1 & \hdots & 0 \\
	t_{1} - x_{m} & 1 & 0 &  \hdots & 0 & 0 & \hdots & 0 \\
	t_{2} - x_{m} & 0 & 1 &  \hdots & 0 & 0 & \hdots & 0 \\
	\vdots & \vdots & \vdots & \ddots & \vdots  & \vdots & \ddots  & \vdots \\
	t_{n} - x_{m} & 0 & 0 &  \hdots & 1 & 0 & \hdots & 0 \\
}
\in \mathbb{R}^{nm \times (n+m-1)},
\]
\[
\boldsymbol{\gamma}^{(3)}_{\boldsymbol{t-x}} = 
\begin{bmatrix}
\gamma^{(3)}_{t_{1}-x_{m}}\\
\gamma^{(3)}_{t_{2}-x_{m}}\\
\vdots\\
\gamma^{(3)}_{t_{n}-x_{m}}\\
\gamma^{(3)}_{t_{n}-x_{m-1}}\\
\vdots\\
\gamma^{(3)}_{t_{n}-x_{1}}\\
\end{bmatrix}
\in \mathbb{R}^{(n+m-1)\times 1},
\text{and} \hspace{0.2cm}
\boldsymbol{\epsilon}=
\begin{bmatrix} 
\boldsymbol{\epsilon}_{{x}_{1}}\\ \boldsymbol{\epsilon}_{{x}_{2}} \\ \vdots \\ \boldsymbol{\epsilon}_{{x}_{m}}
\end{bmatrix}
\in \mathbb{R}^{nm\times 1}.
\]
\par In aid of explaining the above full matrix representation of the proposed model, here we consider a smaller matrix of cells where the row and column of the matrix directly correspond to year $t$ and age $x$ without the use of script for illustration purposes. The cohort effects parameter $\gamma_{t-x}^{(3)}$ is indexed by the year-of-birth and its corresponding value on each cell is displayed in Table \ref{table: Values of the cohort index} for the cohort effects. We can see that the index values of $\gamma_{t-x}^{(3)}$ are the same in the `cohort direction', which are on the cells $(x, t)$, $(x + 1, t + 1)$ and so on. The total number of cohort effects parameters is equal to the number of rows plus the number of columns minus one, which is four in our illustration here. The proposed model with the smaller size of matrix cells can be expressed as
\[
\begin{bmatrix} 
{y}_{\textcolor{black}{x_{1},t_{1}}} \\ {y}_{\textcolor{black}{x_{1},t_{2}}} \\ {y}_{\textcolor{black}{x_{2},t_{1}}} \\ {y}_{\textcolor{black}{x_{2},t_{2}}} \\ {y}_{\textcolor{black}{x_{3},t_{1}}} \\ {y}_{\textcolor{black}{x_{3},t_{2}}}
\end{bmatrix}
=
\begin{bmatrix} 
1 & t_{1} \\ 1 & t_{2} \\ 1 & t_{1}  \\ 1 & t_{2}  \\  1 & t_{1}\\ 1 & t_{2}
\end{bmatrix}
\begin{bmatrix} 
\beta^{(1)}\\
\beta^{(2)}\\
\end{bmatrix}
+
\begin{bmatrix} 
1 & 0 & 0 \\ 1 & 0 & 0  \\ 0 & 1 & 0 \\ 0 & 1 & 0  \\  0 & 0 & 1 \\ 0 & 0 & 1 
\end{bmatrix}
\begin{bmatrix} 
\gamma^{(1)}_{x_{1}}\\
\gamma^{(1)}_{x_{2}}\\
\gamma^{(1)}_{x_{3}}\\
\end{bmatrix}
+
\begin{bmatrix} 
t_{1} & 0 & 0 \\ t_{2} & 0 & 0  \\ 0 & t_{1} & 0 \\ 0 & t_{2} & 0  \\  0 & 0 & t_{1} \\ 0 & 0 & t_{2} 
\end{bmatrix}
\begin{bmatrix} 
\gamma^{(2)}_{x_{1}}\\
\gamma^{(2)}_{x_{2}}\\
\gamma^{(2)}_{x_{3}}\\
\end{bmatrix}
+
\begin{bmatrix} 
0 & 0 & 1 & 0 \\ 0 & 0 & 0 & 1  \\ 0 & 1 & 0 & 0 \\ 0 & 0 & 1 &0  \\ 1& 0 & 0 & 0 \\ 0 & 1 & 0 &0 
\end{bmatrix}
\begin{bmatrix} 
\gamma^{(3)}_{-2}\\
\gamma^{(3)}_{-1}\\
\gamma^{(3)}_{0}\\
\gamma^{(3)}_{1}\\
\end{bmatrix}
+
\begin{bmatrix} 
{\epsilon}_{\textcolor{black}{x_{1},t_{1}}}\\ {\epsilon}_{\textcolor{black}{x_{1},t_{2}}} \\ {\epsilon}_{\textcolor{black}{x_{2},t_{1}}}\\ {\epsilon}_{\textcolor{black}{x_{2},t_{2}}} \\ {\epsilon}_{\textcolor{black}{x_{3},t_{1}}}\\ {\epsilon}_{\textcolor{black}{x_{3},t_{2}}}
\end{bmatrix}.
\]

\begin{table}
	\begin{tabularx}{1\textwidth} { 
			| >{\centering\arraybackslash}X 
			| >{\centering\arraybackslash}X
			| >{\centering\arraybackslash}X  
			| >{\centering\arraybackslash}X | }
		\hline
		year $t$ $/$ age $x$ & $x = 1$ & $x = 2$ & $x =3$ \\
		\hline
		$t = 1$  & $\gamma_{0}^{(3)}$  & $\gamma_{-1}^{(3)}$  & $\gamma_{-2}^{(3)}$ \\
		\hline
		$t = 2$  & $\gamma_{1}^{(3)}$ & $\gamma_{0}^{(3)}$   & $\gamma_{-1}^{(3)}$  \\
		\hline
	\end{tabularx}
	\caption{Values of the cohort effects parameters $\gamma_{t-x}^{(3)}$ in a matrix of cells $(t, x)$.}
	\label{table: Values of the cohort index}
\end{table}

For the stochastic components in the proposed model, we assume that the prior distributions of $\boldsymbol{\gamma}^{(1)}_{\boldsymbol{x}} \sim \textcolor{black}{\mathcal{N}(\boldsymbol{0},\boldsymbol{K}^{(1)})}$, $\boldsymbol{\gamma}^{(2)}_{\boldsymbol{x}}\sim \textcolor{black}{\mathcal{N}(\boldsymbol{0},\boldsymbol{K}^{(2)})}$ and $\boldsymbol{\gamma}^{(3)}_{\boldsymbol{t-x}}\sim \textcolor{black}{\mathcal{N}(\boldsymbol{0},\boldsymbol{K}^{(3)})}$. \textcolor{black}{$\boldsymbol{K}^{(1)}$} and \textcolor{black}{$\boldsymbol{K}^{(2)}$} are both $m \times m$ squared exponential covariance matrices which are used to model the dependence structure among different age groups. They have the following form, for $i = 1,2$, i.e.
\[\textcolor{black}{\boldsymbol{K}^{(i)}}= \textcolor{black}{h_{i}^{2}}\text{exp} \Bigg(-
\begin{bmatrix} 
\frac{(x_{1}-x_{1})^{2}}{2\textcolor{black}{l_{i}}} & \frac{(x_{1}-x_{2})^{2}}{2\textcolor{black}{l_{i}}} & \hdots & \frac{(x_{1}-x_{m})^{2}}{2\textcolor{black}{l_{i}}} \\ \frac{(x_{2}-x_{1})^{2}}{2\textcolor{black}{l_{i}}} & \frac{(x_{2}-x_{2})^{2}}{2\textcolor{black}{l_{i}}}  & \hdots & \frac{(x_{2}-x_{m})^{2}}{2\textcolor{black}{l_{i}}} \\ \vdots & \vdots & \ddots & \vdots \\ \frac{(x_{m}-x_{1})^{2}}{2\textcolor{black}{l_{i}}} &  \frac{(x_{m}-x_{2})^{2}}{2\textcolor{black}{l_{i}}} & \hdots & \frac{(x_{m}-x_{m})^{2}}{2\textcolor{black}{l_{i}}}
\end{bmatrix} 
\Bigg)
,
\]
where \textcolor{black}{$h_{i}$} is the response-scale amplitude parameter determining the variation of function values \textcolor{black}{, and} \textcolor{black}{$l_{i}$} is the characteristic length-scale parameter that gives smooth variations in a covariate-scale and controls how close the varying age groups should correlate. 
\par \textcolor{black}{$\boldsymbol{K}^{(3)}$} is another $(n+m-1) \times (n+m-1)$ squared exponential covariance matrix for  $\boldsymbol{\gamma}^{(3)}_{\boldsymbol{t-x}}$ modelling the dependence structure among different groups of people born in the same year with the following form
\[\textcolor{black}{\boldsymbol{K}^{(3)}} = c^{2}\text{exp}\Bigg(-
\begin{bmatrix} 
	\frac{((t_{1} - x_{m})-(t_{1} - x_{m}))^{2}}{2s} & \frac{((t_{1} - x_{m})-(t_{2} - x_{m}))^{2}}{2s} & \hdots & \frac{((t_{1} - x_{m})-(t_{n} - x_{1}))^{2}}{2s} \\ \frac{((t_{2} - x_{m})-(t_{1} - x_{m}))^{2}}{2s} & \frac{((t_{2} - x_{m})-(t_{2} - x_{m}))^{2}}{2s} & \hdots & \frac{((t_{2} - x_{m})-(t_{n} - x_{1}))^{2}}{2s} \\ \vdots & \vdots & \ddots & \vdots \\ \frac{((t_{n} - x_{1})-(t_{1} - x_{m}))^{2}}{2s} &  \frac{((t_{n} - x_{1})-(t_{2} - x_{m}))^{2}}{2s}  & \hdots & \frac{((t_{n} - x_{1})-(t_{n} - x_{1}))^{2}}{2s} 
\end{bmatrix}
\Bigg)
,
\]
where $c$ is the response-scale amplitude parameter that determines the variation of function values\textcolor{black}{, and $s$} is the characteristic length-scale parameter that gives smooth variations in a covariate-scale and controls how close the different groups of people born in the same year should correlate.  
\par The main justification of choosing the squared exponential covariance matrices here for all the stochastic components is to reflect the dependencies among different age groups properly as we assume that people who are closer in age groups or birth year would tend to have more similarities in their mortality patterns than those who are in \textcolor{black}{distant} age ranges or were born in a different generation.
\par With the prior distribution assumptions among all the random effects components, Equation (\ref{eq: (matrix) age group specific deviation trend function}) can be restated as a multivariate normal distribution among all its random effects components, i.e.
\begin{equation} \label{eq: (matrix with prior) age group specific deviation trend function}
\boldsymbol{Y} = \boldsymbol{T}\boldsymbol{\beta} + \boldsymbol{Z}^{(1)}\boldsymbol{\gamma}^{(1)}_{\boldsymbol{x}} +  \boldsymbol{Z}^{(2)}\boldsymbol{\gamma}^{(2)}_{\boldsymbol{x}} +  \boldsymbol{Z}^{(3)}\boldsymbol{\gamma}^{(3)}_{\boldsymbol{t-x}} + \boldsymbol{\epsilon},
\end{equation}
where
\[
\begin{bmatrix}
\boldsymbol{\gamma}^{(1)}_{\boldsymbol{x}}\\
\boldsymbol{\gamma}^{(2)}_{\boldsymbol{x}}\\
\boldsymbol{\gamma}^{(3)}_{\boldsymbol{t-x}}\\
\boldsymbol{\varepsilon}
\end{bmatrix}
\sim
\mathcal{N} \Bigg(
\begin{bmatrix}
\boldsymbol{0} \\
\boldsymbol{0} \\
\boldsymbol{0} \\
\boldsymbol{0}
\end{bmatrix}
,
\begin{bmatrix}
	\textcolor{black}{\boldsymbol{K}^{(1)}} & \boldsymbol{0} & \boldsymbol{0} & \boldsymbol{0} \\
	\boldsymbol{0}& \textcolor{black}{\boldsymbol{K}^{(2)}} & \boldsymbol{0} & \boldsymbol{0}\\
	\boldsymbol{0} & \boldsymbol{0}& \textcolor{black}{\boldsymbol{K}^{(3)}} & \boldsymbol{0}\\
	\boldsymbol{0} & \boldsymbol{0} & \boldsymbol{0} & \sigma^{2}\boldsymbol{I} \\
\end{bmatrix}
\bigg). 
\]
\subsubsection{Model parameters inference}
With the model formulation specified in the previous section, we can infer the model parameters by the Bayesian paradigm. The probability distribution of $\boldsymbol{Y}$ in Equation (\ref{eq: (matrix with prior) age group specific deviation trend function}) is given by $\mathcal{N}(\boldsymbol{T}\boldsymbol{\beta}, \boldsymbol{V(\theta)})$ where $\boldsymbol{V(\theta)} = \boldsymbol{Z}^{(1)}\textcolor{black}{\boldsymbol{K}^{(1)}}{\boldsymbol{Z}^{(1)}}^{T} + \boldsymbol{Z}^{(2)}\textcolor{black}{\boldsymbol{K}^{(2)}}{\boldsymbol{Z}^{(2)}}^{T} + \boldsymbol{Z}^{(3)}\textcolor{black}{\boldsymbol{K}^{(3)}}{\boldsymbol{Z}^{(3)}}^{T} + \sigma^{2}\boldsymbol{I}$. $\boldsymbol{V(\theta)}$ is the $nm \times nm$ variance-covariance matrix parametrised by the (hyper-) parameters $\boldsymbol{\theta} = [h^{(1)}, l^{(1)},h^{(2)}, l^{(2)},c,s]^{T}$ involved in its components covariance matrices $\textcolor{black}{\boldsymbol{K}^{(1)}}$, $\textcolor{black}{\boldsymbol{K}^{(2)}}$, $\textcolor{black}{\boldsymbol{K}^{(3)}}$ and the random \textcolor{black}{error} variance $\sigma^{2}$. Hence the log-likelihood function of $\boldsymbol{\beta}$, the parameters of the variance-covariance matrix  $\boldsymbol{\theta}$ and the random \textcolor{black}{error} variance $\sigma^{2}$ is
\begin{equation} \label{eq: random effect marginal likelihood function}
	\mathcal{L}(\boldsymbol{\beta}, \boldsymbol{\theta},\sigma^{2}) =  -\frac{1}{2}\text{log}\vert\boldsymbol{V(\theta)}\vert -\frac{1}{2}(\boldsymbol{Y} - \boldsymbol{T}\boldsymbol{\beta})^{T}\boldsymbol{V(\theta)}^{-1}(\boldsymbol{Y} - \boldsymbol{T}\boldsymbol{\beta}) - \frac{nm}{2}\text{log}(2\pi),
\end{equation} 
where $|\cdot|$ is the determinant of a matrix.  
\par 
Given that there is no analytical closed-form solution to this maximisation problem, we need to resort to some standard gradient-based numerical optimisation techniques, such as the Newton-Raphson method or the Conjugate Gradient method, to maximise the log-likelihood function $\mathcal{L}(\boldsymbol{\beta}, \boldsymbol{\theta},\sigma^{2})$ in Equation (\ref{eq: random effect marginal likelihood function}) for obtaining the optimal estimates of the model parameters.
\par To set the form of the model parameters appropriate for the iterative process of the gradient-based optimisation method, we seek to differentiate the log-likelihood function $\mathcal{L}(\boldsymbol{\beta}, \boldsymbol{\theta},\sigma^{2})$ with respect to its elements $\boldsymbol{\beta}$, $\boldsymbol{\theta}$, and $\sigma^{2}$. This yields the following partial derivatives
\begin{equation}\label{eq: partial  partial derivative of beta}
	\frac{\partial\mathcal{L}(\boldsymbol{\beta}, \boldsymbol{\theta},\sigma^{2})}{\partial \boldsymbol{\beta}} = -(\boldsymbol{T}^{T}\boldsymbol{V(\theta)}^{-1}\boldsymbol{T}\boldsymbol{\beta} - \boldsymbol{T}^{T}\boldsymbol{V(\theta)}^{-1}\boldsymbol{Y}),
\end{equation}
\[
\begin{aligned}
	\frac{\partial\mathcal{L}(\boldsymbol{\beta}, \boldsymbol{\theta},\sigma^{2})}{\partial  \theta_{i}} &= -\dfrac{1}{2}\bigg\lbrace  \text{tr}\bigg( \boldsymbol{V(\theta)}^{-1}\frac{\partial \boldsymbol{V(\theta)}}{\partial  \theta_{i}} \bigg) + (\boldsymbol{Y} - \boldsymbol{T}\boldsymbol{\beta})^{T}\bigg(\boldsymbol{V(\theta)}^{-1}\frac{\partial \boldsymbol{V(\theta)}}{\partial  \theta_{i}}\boldsymbol{V(\theta)}^{-1}\bigg)(\boldsymbol{Y} - \boldsymbol{T}\boldsymbol{\beta}) \bigg\rbrace,
\end{aligned}
\]
\[
\frac{\partial\mathcal{L}(\boldsymbol{\beta}, \boldsymbol{\theta},\sigma^{2})}{\partial  \sigma^{2}} = -\dfrac{1}{2}\bigg\lbrace \text{tr}(\boldsymbol{V(\theta)}^{-1}) + (\boldsymbol{Y} - \boldsymbol{T}\boldsymbol{\beta})^{T}(\boldsymbol{V(\theta)}^{-1}\boldsymbol{V(\theta)}^{-1})(\boldsymbol{Y} - \boldsymbol{T}\boldsymbol{\beta})\bigg\rbrace,
\]
where $\theta_{i}$ is a generic notation of the $i^{\text{th}}$ element of parameters in the vector $\boldsymbol{\theta}$.
\par Setting Equation (\ref{eq: partial  partial derivative of beta}) equal to zero and solving it give us the closed-form maximum likelihood estimator $\hat{\boldsymbol{\beta}}$, i.e.
\begin{equation}\label{eq: reduced partial derivative of beta}
	\hat{\boldsymbol{\beta}} =(\boldsymbol{T}^{T}\boldsymbol{V(\theta)}^{-1}\boldsymbol{T})^{-1} \boldsymbol{T}^{T}\boldsymbol{V(\theta)}^{-1}\boldsymbol{Y}.
\end{equation}
The typical iterative procedure in the gradient-based optimiser for estimating ${\textcolor{black}{\boldsymbol{\beta}}}$, $\boldsymbol{\theta}$ and $\sigma^{2}$ with the above partial derivatives involves the following four steps: 
\begin{enumerate}
	\item[1.] Assign initial values to ${\boldsymbol{\hat{\theta}}}_{(1)} = [\hat{h}_{(1)}^{(1)}, \hat{l}_{(1)}^{(1)},\hat{h}_{(1)}^{(2)}, \hat{l}_{(1)}^{(2)},\hat{c}_{(1)},\hat{s}_{(1)}]^{T}$ and $\hat{\sigma}^{2}_{(1)}$.
	\item[2.] Substitute $\boldsymbol{\hat{\theta}}_{(k)}$ and $\hat{\sigma}^{2}_{(k)}$ to solve for $\hat{\boldsymbol{\beta}}_{(k)}$ in Equation (\ref{eq: reduced partial derivative of beta}) and \textcolor{black}{calculate the value of the following log-likelihood function in each $k^{\text{th}}$ step iteration:}
	\[
	\begin{aligned}
		\mathcal{L}(\boldsymbol{\hat{\beta}}_{(k)}, \boldsymbol{\hat{\theta}}_{(k)},\hat{\sigma}^{2}_{(k)}) =  -\frac{1}{2}\text{log}\vert\boldsymbol{V}(\boldsymbol{\hat{\theta}}_{(k)})\vert -\frac{1}{2}(\boldsymbol{Y} - \boldsymbol{T}\boldsymbol{\hat{\beta}}_{(k)})^{T}\boldsymbol{V}(\boldsymbol{\hat{\theta}}_{(k)})^{-1}(\boldsymbol{Y} - \boldsymbol{T}\boldsymbol{\hat{\beta}}_{(k)}) -\frac{nm}{2}\text{log}(2\pi).
	\end{aligned}
	\]   
	\item[3.] Update ${\hat{\theta}}_{i,(k+1)} = {\hat{\theta}}_{i,(k)} + \frac{{\partial\mathcal{L}(\boldsymbol{\hat{\beta}}_{(k)}, \boldsymbol{\hat{\theta}}_{(k)},\hat{\sigma}^{2}_{(k)})/\partial \hat{\theta}_{i,(k)}}} {{\partial^{2}\mathcal{L}(\boldsymbol{\hat{\beta}}_{(k)}, \boldsymbol{\hat{\theta}}_{(k)},\hat{\sigma}^{2}_{(k)})/\partial^{2}  \hat{\theta}_{i,(k)}}}$ and $\hat{\sigma}^{2}_{(k+1)} = \hat{\sigma}^{2}_{(k)} + \frac{{\partial\mathcal{L}(\boldsymbol{\hat{\beta}}_{(k)}, \boldsymbol{\hat{\theta}}_{(k)},\hat{\sigma}^{2}_{(k)})/\partial \hat{\sigma}^{2}_{(k)}} }{\partial^{2}\mathcal{L}(\boldsymbol{\hat{\beta}}_{(k)}, \boldsymbol{\hat{\theta}}_{(k)},\hat{\sigma}^{2}_{(k)})/\partial^{2}  \hat{\sigma}^{2}_{(k)}}$ based on the gradient ascent criteria.
	\item[4.] Repeat Step 2 and Step 3 until \textcolor{black}{the above log-likelihood function converges} \footnote{The performances of the gradient-based optimisation approaches may depend on the initial values supplied. Several optimisations running can be applied with different initial values may be necessary if no prior knowledge about the function to optimise.}.
\end{enumerate}
Once the estimated $\boldsymbol{\hat{\theta}}_{(p)}$, $\hat{\sigma}^{2}_{(p)}$ \textcolor{black}{and $\boldsymbol{V}(\boldsymbol{\hat{\theta}}_{(p)})$ are obtained in the $p^{\text{th}}$ step iteration where the log-likelihood function converges, we have}
{
	\[
	\begin{bmatrix}
		\boldsymbol{Y}\\
		\boldsymbol{\gamma}^{(1)}_{\boldsymbol{x}}\\
		\boldsymbol{\gamma}^{(2)}_{\boldsymbol{x}}\\
		\boldsymbol{\gamma}^{(3)}_{\boldsymbol{t-x}}\\
	\end{bmatrix}
	\sim
	\mathcal{N} \Bigg(
	\begin{bmatrix}
		\boldsymbol{T}\hat{\boldsymbol{\beta}}_{(p)}\\
		\boldsymbol{0} \\
		\boldsymbol{0} \\
		\boldsymbol{0}
	\end{bmatrix}
	,
	\begin{bmatrix}
		\boldsymbol{V}(\boldsymbol{\hat{\theta}}_{(p)})  & \boldsymbol{Z}^{(1)}\textcolor{black}{\hat{\boldsymbol{K}}^{(1)}} & \boldsymbol{Z}^{(2)}\textcolor{black}{\hat{\boldsymbol{K}}^{(2)}} & \boldsymbol{Z}^{(3)}\textcolor{black}{\hat{\boldsymbol{K}}^{(3)}}\\
		\textcolor{black}{\hat{\boldsymbol{K}}^{(1)}}{\boldsymbol{Z}^{(1)}}^{T} & \textcolor{black}{\hat{\boldsymbol{K}}^{(1)}} & \boldsymbol{0} & \boldsymbol{0}\\
		\textcolor{black}{\hat{\boldsymbol{K}}^{(2)}}{\boldsymbol{Z}^{(2)}}^{T} & \boldsymbol{0}& \textcolor{black}{\hat{\boldsymbol{K}}^{(2)}} & \boldsymbol{0}\\
		\textcolor{black}{\hat{\boldsymbol{K}}^{(3)}}{\boldsymbol{Z}^{(3)}}^{T}  & \boldsymbol{0} & \boldsymbol{0} & \textcolor{black}{\hat{\boldsymbol{K}}^{(3)}} \\
	\end{bmatrix}
	\Bigg),
	\]
}and the conditional distributions of  $\boldsymbol{\gamma}^{(1)}_{\boldsymbol{x}}$, $\boldsymbol{\gamma}^{(2)}_{\boldsymbol{x}}$, $\boldsymbol{\gamma}^{(3)}_{\boldsymbol{t-x}}$ given $\boldsymbol{Y}$ follow the multivariate Gaussian distribution, which are $\boldsymbol{\gamma}^{(1)}_{\boldsymbol{x}}\vert\boldsymbol{Y} \sim \mathcal{N}(\boldsymbol{\hat{\gamma}}^{(1)}_{\boldsymbol{x}}, \text{Var}(\boldsymbol{{\gamma}}^{(1)}_{\boldsymbol{x}}))$, $\boldsymbol{\gamma}^{(2)}_{\boldsymbol{x}}\vert\boldsymbol{Y} \sim \mathcal{N}(\boldsymbol{\hat{\gamma}}^{(2)}_{\boldsymbol{x}}, \text{Var}(\boldsymbol{{\gamma}}^{(2)}_{\boldsymbol{x}}))$ and $\boldsymbol{\gamma}^{(3)}_{\boldsymbol{t-x}}\vert\boldsymbol{Y} \sim \mathcal{N}(\boldsymbol{\hat{\gamma}}^{(3)}_{\boldsymbol{t-x}}, \text{Var}(\boldsymbol{{\gamma}}^{(3)}_{\boldsymbol{t-x}}))$, \textcolor{black}{where}
\begin{equation}\label{eq: conditional mean of estimated gamma_1}
	\boldsymbol{\hat{\gamma}}^{(1)}_{\boldsymbol{x}}= \textcolor{black}{\hat{\boldsymbol{K}}^{(1)}}{\boldsymbol{Z}^{(1)}}^{T}\boldsymbol{V}(\boldsymbol{\hat{\theta}}_{(p)})^{-1}(\boldsymbol{Y} - \boldsymbol{T}\hat{\boldsymbol{\beta}}_{(p)}),
\end{equation}
\[
\text{Var}(\boldsymbol{{\gamma}}^{(1)}_{\boldsymbol{x}}) = \textcolor{black}{\hat{\boldsymbol{K}}^{(1)}} - [  \textcolor{black}{\hat{\boldsymbol{K}}^{(1)}}{\boldsymbol{Z}^{(1)}}^{T}\boldsymbol{V}(\boldsymbol{\hat{\theta}}_{(p)})^{-1}{\boldsymbol{Z}^{(1)}}\textcolor{black}{\hat{\boldsymbol{K}}^{(1)}}] ,
\]
\begin{equation}\label{eq: conditional mean of estimated gamma_2}
	\boldsymbol{{\hat{\gamma}}}^{(2)}_{\boldsymbol{x}} = \textcolor{black}{\hat{\boldsymbol{K}}^{(2)}}{\boldsymbol{Z}^{(2)}}^{T}\boldsymbol{V}(\boldsymbol{\hat{\theta}}_{(p)})^{-1}(\boldsymbol{Y} - \boldsymbol{T}\hat{\boldsymbol{\beta}}_{(p)}),
\end{equation}
\[
\text{Var}(\boldsymbol{{\gamma}}^{(2)}_{\boldsymbol{x}}) = \textcolor{black}{\hat{\boldsymbol{K}}^{(2)}} - [\textcolor{black}{\hat{\boldsymbol{K}}^{(2)}}{\boldsymbol{Z}^{(2)}}^{T}\boldsymbol{V}(\boldsymbol{\hat{\theta}}_{(p)})^{-1}{\boldsymbol{Z}^{(2)}}\textcolor{black}{\hat{\boldsymbol{K}}^{(2)}}],
\]
\begin{equation}\label{eq: conditional mean of estimated gamma_3}
	\boldsymbol{\hat{\gamma}}^{(3)}_{\boldsymbol{t-x}} = \textcolor{black}{\hat{\boldsymbol{K}}^{(3)}}{\boldsymbol{Z}^{(3)}}^{T}\boldsymbol{V}(\boldsymbol{\hat{\theta}}_{(p)})^{-1}(\boldsymbol{Y} - \boldsymbol{T}\hat{\boldsymbol{\beta}}_{(p)}),
\end{equation}
\[
\text{Var}(\boldsymbol{{\gamma}}^{(3)}_{\boldsymbol{t-x}}) = \textcolor{black}{\hat{\boldsymbol{K}}^{(3)}} - [\textcolor{black}{\hat{\boldsymbol{K}}^{(3)}}{\boldsymbol{Z}^{(3)}}^{T}\boldsymbol{V}(\boldsymbol{\hat{\theta}}_{(p)})^{-1}{\boldsymbol{Z}^{(3)}}\textcolor{black}{\hat{\boldsymbol{K}}^{(3)}}].
\]
The linear fixed effects estimator $\hat{\boldsymbol{\beta}}$ and its variance $\text{Var}({\boldsymbol{\beta}})$ in the proposed model is
\[
\hat{\boldsymbol{\beta}} = (\boldsymbol{T}^{T}\boldsymbol{V}(\boldsymbol{\hat{\theta}}_{(p)})^{-1}\boldsymbol{T})^{-1} \boldsymbol{T}^{T}\boldsymbol{V}(\boldsymbol{\hat{\theta}}_{(p)})^{-1}\boldsymbol{Y}, 
\]
\[
\text{Var}({\boldsymbol{\beta}}) = \hat{\sigma}^{2}_{(p)}\boldsymbol{I}(\boldsymbol{T}^{T}\boldsymbol{V}(\boldsymbol{\hat{\theta}}_{(p)})^{-1}\boldsymbol{T})^{-1}.
\]
We lastly combine all the estimated components to obtain the complete matrix form of the estimated mean of the proposed model, i.e.
\[
\hat{\boldsymbol{Y}} = \boldsymbol{T}\hat{\boldsymbol{\beta}} + {\boldsymbol{Z}^{(1)}}\boldsymbol{\hat{\gamma}}^{(1)}_{\boldsymbol{x}} +  {\boldsymbol{Z}^{(2)}}\boldsymbol{\hat{\gamma}}^{(2)}_{\boldsymbol{x}}+  {\boldsymbol{Z}^{(3)}}\boldsymbol{\hat{\gamma}}^{(3)}_{\boldsymbol{t-x}},
\]  
and the variance of the estimated mean in the proposed model is
\[
\text{Var}({\boldsymbol{Y}}) =  \boldsymbol{T}\text{Var}({\boldsymbol{\beta}})\boldsymbol{T}^{T}+ \boldsymbol{Z}^{(1)}\text{Var}(\boldsymbol{{\gamma}}^{(1)}_{\boldsymbol{x}}){\boldsymbol{Z}^{(1)}}^{T} + \boldsymbol{Z}^{(2)}\text{Var}(\boldsymbol{{\gamma}}^{(2)}_{\boldsymbol{x}}){\boldsymbol{Z}^{(2)}}^{T} + \boldsymbol{Z}^{(3)}\text{Var}(\boldsymbol{{\gamma}}^{(3)}_{\boldsymbol{t-x}}){\boldsymbol{Z}^{(3)}}^{T} + \hat{\sigma}^{2}_{(p)}\boldsymbol{I}.
\]
\subsubsection{Out-of-sample forecasts and the prediction intervals}
In \textcolor{black}{the proposed model}, we can extrapolate a $h$-step forecast of the logit of the initial mortality rate from the last observed calendar year $t_{n}$ to the calendar year $t_{n+h}$ by extending the row of \textcolor{black}{covariate} matrices ${\boldsymbol{T}}$, ${\boldsymbol{Z}^{(1)}}$, ${\boldsymbol{Z}^{(2)}}$ from the size of $n \times m$ to $(n+h)\times m$ with $[1_{n+1},1_{n+2},\ldots,1_{n+h}]^{T}$, new input $[t_{n+1},t_{n+2},\ldots,t_{n+h}]^{T}$ and reconstructing ${\boldsymbol{Z}^{(3)}}$ as a new $((n+h)\times m) \times (n+h+m-1)$ cohort pattern dummy \textcolor{black}{covariate} matrix in Equation (\ref{eq: (matrix) age group specific deviation trend function}) denoted by ${\boldsymbol{T}^{*}}$, ${\boldsymbol{Z}^{*(1)}}$, ${\boldsymbol{Z}^{*(2)}}$, ${\boldsymbol{Z}^{*(3)}}$, such that
\[
\hat{\boldsymbol{Y}}^{*} = \boldsymbol{T}^{*}\hat{\boldsymbol{\beta}} + {\boldsymbol{Z}^{*(1)}}\boldsymbol{\hat{\gamma}}^{(1)}_{\boldsymbol{x}} +  {\boldsymbol{Z}^{*(2)}}\boldsymbol{\hat{\gamma}}^{(2)}_{\boldsymbol{x}}+  {\boldsymbol{Z}^{*(3)}}\boldsymbol{\hat{\gamma}}^{(3)}_{\boldsymbol{t+h-x}},
\] 
where
\begin{equation} \label{eq: extrapolation of the gamma_3}
	\boldsymbol{\hat{\gamma}}^{(3)}_{\boldsymbol{t+h-x}} = \textcolor{black}{\hat{\boldsymbol{K}}^{*(3)}}{\boldsymbol{Z}^{(3)}}^{T}\boldsymbol{V}(\boldsymbol{\hat{\theta}}_{(p)})^{-1}(\boldsymbol{Y} - \boldsymbol{T}\boldsymbol{\hat{\beta}}),
\end{equation}
\textcolor{black}{and $\hat{\boldsymbol{K}}^{*(3)}$ is the size of $(n+h+m-1) \times (n+m-1)$ covariance matrix for extrapolating $\boldsymbol{\hat{\gamma}}^{(3)}_{\boldsymbol{t-x}}$ to $h$-step ahead , i.e.}
\[\textcolor{black}{\hat{\boldsymbol{K}}^{*(3)}}= \textcolor{black}{\hat{c}_{(p)}^{2}}\text{exp}\Bigg(- 
\begin{bmatrix} 
	\frac{((t_{1} - x_{m})-(t_{1} - x_{m}))^{2}}{2\textcolor{black}{\hat{s}_{(p)}}} & \frac{((t_{1} - x_{m})-(t_{2} - x_{m}))^{2}}{2\textcolor{black}{\hat{s}_{(p)}}} & \hdots & \frac{((t_{1} - x_{m})-(t_{n} - x_{1}))^{2}}{2\textcolor{black}{\hat{s}_{(p)}}} \\ \frac{((t_{2} - x_{m})-(t_{1} - x_{m}))^{2}}{2\textcolor{black}{\hat{s}_{(p)}}} & \frac{((t_{2} - x_{m})-(t_{2} - x_{m}))^{2}}{2\textcolor{black}{\hat{s}_{(p)}}} & \hdots & \frac{((t_{2} - x_{m})-(t_{n} - x_{1}))^{2}}{2\textcolor{black}{\hat{s}_{(p)}}} \\ \vdots & \vdots & \ddots & \vdots \\ \frac{((\textcolor{black}{t_{n+h}} - x_{1})-(t_{1} - x_{m}))^{2}}{2\textcolor{black}{\hat{s}_{(p)}}} &  \frac{((\textcolor{black}{t_{n+h}} - x_{1})-(t_{2} - x_{m}))^{2}}{2\textcolor{black}{\hat{s}_{(p)}}}  & \hdots & \frac{((\textcolor{black}{t_{n+h}} - x_{1})-(t_{n} - x_{1}))^{2}}{2\textcolor{black}{\hat{s}_{(p)}}}
\end{bmatrix}
\Bigg).
\]

The $h$-step out-of-sample forecast variance \textcolor{black}{by} the proposed model is
{
	\[
	\text{Var}({{\boldsymbol{Y}}^{*}}) = \boldsymbol{T}^{*}\text{Var}({\boldsymbol{\beta}}){\boldsymbol{T}^{*}}^{T} + {\boldsymbol{Z}^{*(1)}}\text{Var}(\boldsymbol{{\gamma}}^{(1)}_{\boldsymbol{x}}){\boldsymbol{Z}^{*(1)}}^{T} + {\boldsymbol{Z}^{*(2)}}\text{Var}(\boldsymbol{{\gamma}}^{(2)}_{\boldsymbol{x}}){\boldsymbol{Z}^{*(2)}}^{T}+  {\boldsymbol{Z}^{*(3)}}\text{Var}(\boldsymbol{{\gamma}}^{(3)}_{\boldsymbol{t+h-x}}){\boldsymbol{Z}^{*(3)}}^{T} + \hat{\sigma}^{2}_{(p)}\boldsymbol{I},
	\]
}
where
\[
\text{Var}(\boldsymbol{{\gamma}}^{(3)}_{\boldsymbol{t+h-x}}) = \textcolor{black}{\hat{\boldsymbol{K}}^{**(3)}} - \textcolor{black}{\hat{\boldsymbol{K}}^{**(3)}}{\boldsymbol{Z}^{(3)}}^{T}\boldsymbol{V}(\boldsymbol{\hat{\theta}}_{(p)})^{-1}{\boldsymbol{Z}^{(3)}}\textcolor{black}{\hat{\boldsymbol{K}}^{**(3)}},
\]
\textcolor{black}{and $\hat{\boldsymbol{K}}^{**(3)}$ is the size of $(n+h+m-1) \times (n+h+m-1)$ covariance matrix, i.e.}
\[\textcolor{black}{\hat{\boldsymbol{K}}^{**(3)}}= \textcolor{black}{\hat{c}_{(p)}^{2}}\text{exp}\Bigg(- 
\begin{bmatrix} 
	\frac{((t_{1} - x_{m})-(t_{1} - x_{m}))^{2}}{2\textcolor{black}{\hat{s}_{(p)}}} & \frac{((t_{1} - x_{m})-(t_{2} - x_{m}))^{2}}{2\textcolor{black}{\hat{s}_{(p)}}} & \hdots & \frac{((t_{1} - x_{m})-(\textcolor{black}{t_{n+h}} - x_{1}))^{2}}{2\textcolor{black}{\hat{s}_{(p)}}} \\ \frac{((t_{2} - x_{m})-(t_{1} - x_{m}))^{2}}{2\textcolor{black}{\hat{s}_{(p)}}} & \frac{((t_{2} - x_{m})-(t_{2} - x_{m}))^{2}}{2\textcolor{black}{\hat{s}_{(p)}}} & \hdots & \frac{((t_{2} - x_{m})-(\textcolor{black}{t_{n+h}} - x_{1}))^{2}}{2\textcolor{black}{\hat{s}_{(p)}}} \\ \vdots & \vdots & \ddots & \vdots \\ \frac{((\textcolor{black}{t_{n+h}} - x_{1})-(t_{1} - x_{m}))^{2}}{2\textcolor{black}{\hat{s}_{(p)}}} &  \frac{((\textcolor{black}{t_{n+h}} - x_{1})-(t_{2} - x_{m}))^{2}}{2\textcolor{black}{\hat{s}_{(p)}}}  & \hdots & \frac{((\textcolor{black}{t_{n+h}} - x_{1})-(\textcolor{black}{t_{n+h}} - x_{1}))^{2}}{2\textcolor{black}{\hat{s}_{(p)}}}
\end{bmatrix}
\Bigg).
\]

\par With the normality assumptions on the model error and the estimated $\text{Var}({{\boldsymbol{Y}}^{*}})$, a $100(1-\alpha)\%$ prediction interval for $\hat{\boldsymbol{Y}}^{*}$ can be calculated as $\hat{\boldsymbol{Y}}^{*}\pm z_{\alpha}(\text{diag}(\text{Var}({{\boldsymbol{Y}}^{*}})))^{1/2}$, where $z_{\alpha}$ is the $(1-\alpha/2)$ quantile of the standard normal distribution.
\section{Applications} \label{Section 4: time-series variant mixed-effects CBD model}
In this section, we demonstrate the usefulness of \textcolor{black}{the proposed model} by modelling and forecasting two sets of sex-specific empirical mortality data. We apply the proposed model to the sex-specific mortality data of Japan for illustration purposes first, then compare and evaluate the quality of the fitted sex-specific mortality curves by the proposed approach with \textcolor{black}{the CBD model} using sex-specific mortality data of ten different developed countries. 
\subsection{Male mortality data} 
The male mortality data of Japan is available from the year 1947 to the year 2016 (70 years in total) from the \cite{human2017university}. The database consists of the age-specific male central mortality rate by a single calendar year of age up to 110 years old. Given \textcolor{black}{that} the CBD model is designed in particular for annuities and pensions schemes for older people, we restrict the numerical examples for the same examined age groups from age 60 to age 89 (30 age groups in total) as \textcolor{black}{the CBD model} \citep{cairns2009quantitative}. \par
\begin{figure}[!thb]
	\centering
	\begin{minipage}{0.78\textwidth}
		\centering
		\includegraphics[width=1\linewidth]{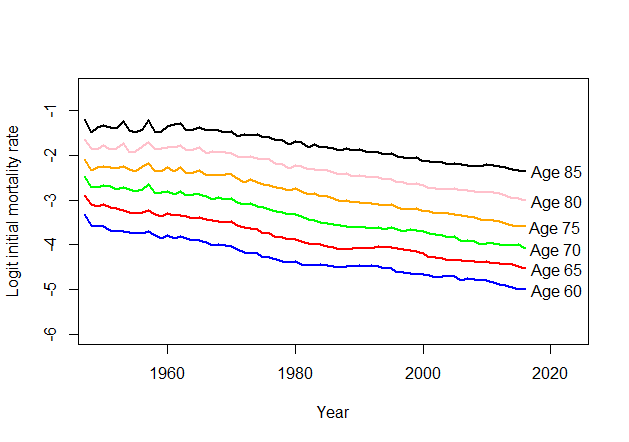}
		\subcaption{}
		\label{fig: Univariate time series of the observed logit initial male mortality rates with 5-year age intervals}
	\end{minipage}
	\begin{minipage}{0.78\textwidth}
		\centering
		\includegraphics[width=1\linewidth]{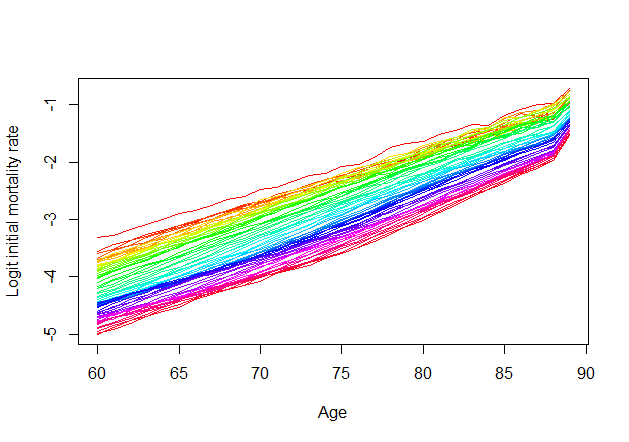}
		\subcaption{}
		\label{fig: Logit male mortality curves}
	\end{minipage}
	\caption{(a) Univariate time series of the observed logit initial male mortality rates with 5-year age intervals and (b) the logit initial male mortality curves from age 60 to age 89 from the year 1947 to the year 2016 in Japan.}
\end{figure}

 We convert the central mortality rate to the initial mortality rate by Equation (\ref{eq: the relationship among q and m}) and present the observed logit initial male mortality rates as univariate time series with 5-year age intervals in Figure \ref{fig: Univariate time series of the observed logit initial male mortality rates with 5-year age intervals} and the observed logit initial male mortality curves from age 60 to age 89 from the year 1947 to the year 2016 in Japan in Figure \ref{fig: Logit male mortality curves}. We can see that there is a general decrease in male mortality rates among all the selected age groups during the examined period. The changing patterns of the time-series male mortality rates look reasonably similar among all the selected groups over the examined period in Figure \ref{fig: Univariate time series of the observed logit initial male mortality rates with 5-year age intervals}. Figure \ref{fig: Logit male mortality curves} presents the same set of data from the age side as a bunch of the mortality curves in a time-ordering indicated by the colours of the rainbow from red to violet. Its vertical pattern shows the general trends and variations of the observed initial male mortality rates across the examined period. We can see that the historical male mortality curves move downwards over time in general, due primarily to the advances in medical technology.
\subsection{Male mortality modelling and forecasting} 
\textcolor{black}{As} the demonstration, we aim to display 10-years-ahead out-of-sample forecasts of the male mortality rates. We first split the dataset into a training dataset with
the observed mortality rates from the year 1947 to the year 2006 and a test dataset with
the remaining observed mortality data from the year 2007 to the last sample data in the year 2016. 
\par
\begin{figure}[!thb]
	\begin{subfigure}{.5\textwidth}
		\centering
		\includegraphics[width=1\linewidth]{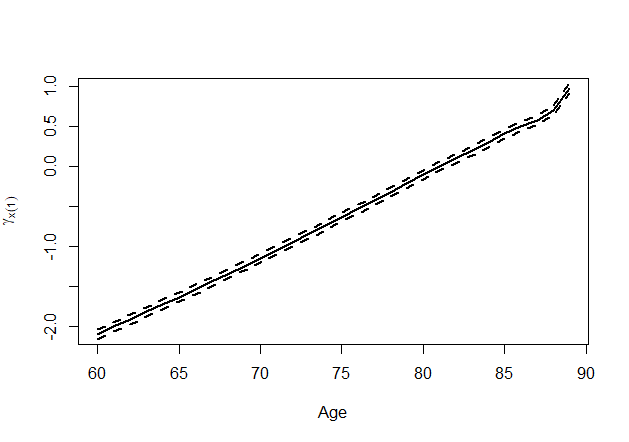}
		\subcaption{}
		\label{fig: Male estimated random effects intercept coefficients}
	\end{subfigure}
	\begin{subfigure}{.5\textwidth}
		\centering
		\includegraphics[width=1\linewidth]{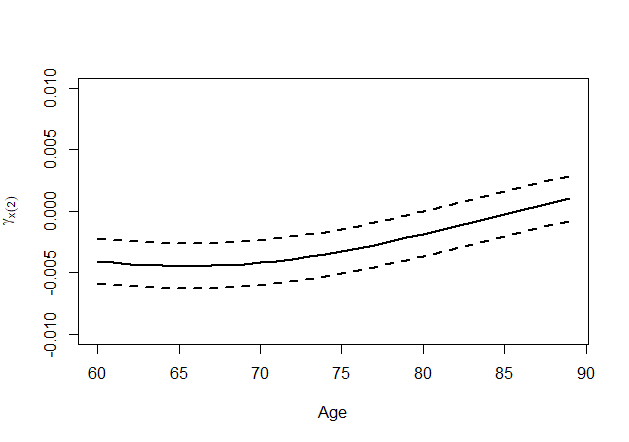}
		\subcaption{}
		\label{fig: Male estimated random effects slope coefficients}
	\end{subfigure} \\[1ex]
	\begin{subfigure}{\linewidth}
		\centering
		\includegraphics[width=0.5\linewidth]{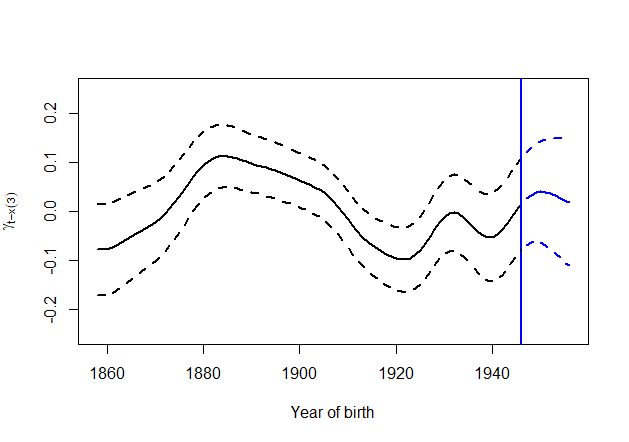}
		\subcaption{}
		\label{fig: Male estimated random cohort effects coefficients}
	\end{subfigure}
	\caption{(a) Estimated random effects intercept coefficients, (b) the estimated random effects slope coefficients and (c) the estimated random cohort effects coefficients using the observed logit initial male mortality rates of Japan from the year 1947 to the year 2006. Dashed lines are the 95\% confidence intervals and the vertical line indicates the start point of the 10-years-ahead forecasts of the estimated random cohort effects coefficients.}
	\label{fig: coefficients of the time-varaint mixed-effects CBD model}
\end{figure}

\begin{figure}[!thb]
	\begin{minipage}{0.5\textwidth}
		\centering
		\includegraphics[width=0.95\linewidth]{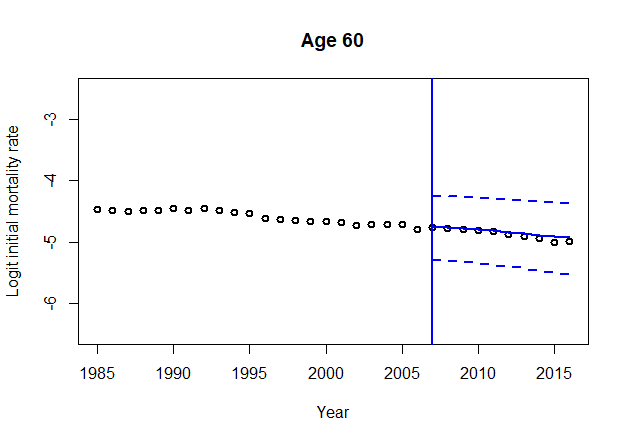}
	\end{minipage}
	\begin{minipage}{0.5\textwidth}
		\centering
		\includegraphics[width=0.95\linewidth]{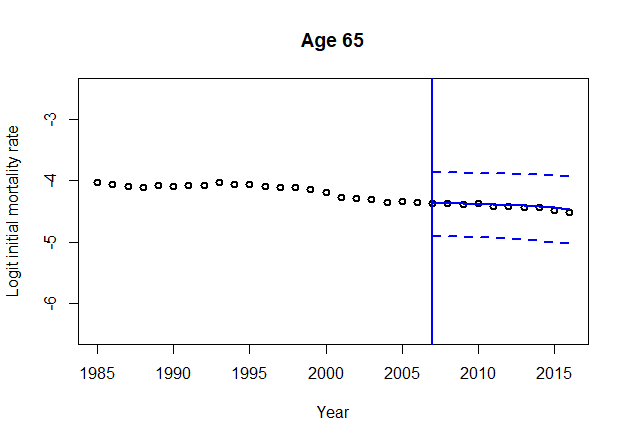}
	\end{minipage}
	\begin{minipage}{0.5\textwidth}
		\centering
		\includegraphics[width=0.95\linewidth]{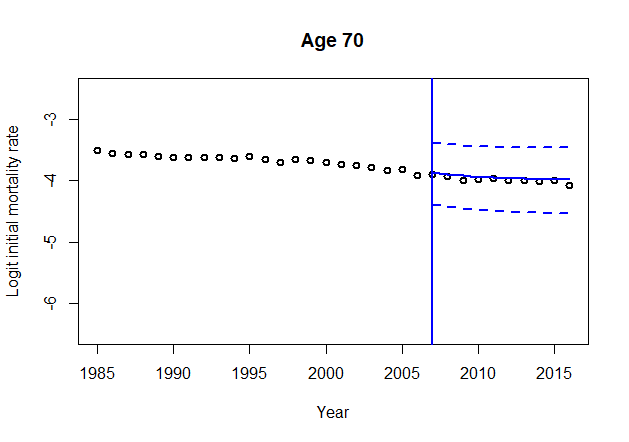}
	\end{minipage} 
	\begin{minipage}{0.5\textwidth}
		\centering
		\includegraphics[width=0.95\linewidth]{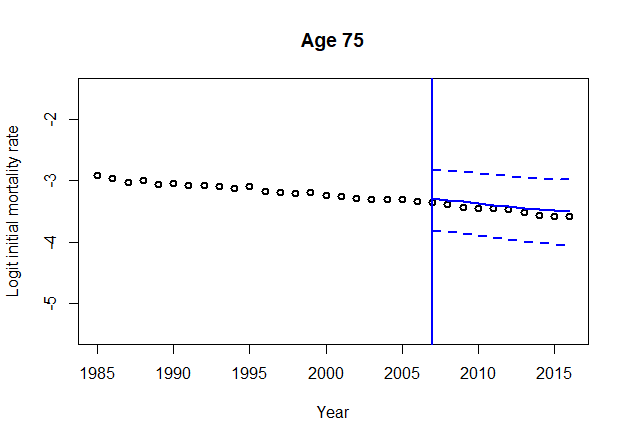}
	\end{minipage}
	\begin{minipage}{0.5\textwidth}
		\centering
		\includegraphics[width=0.95\linewidth]{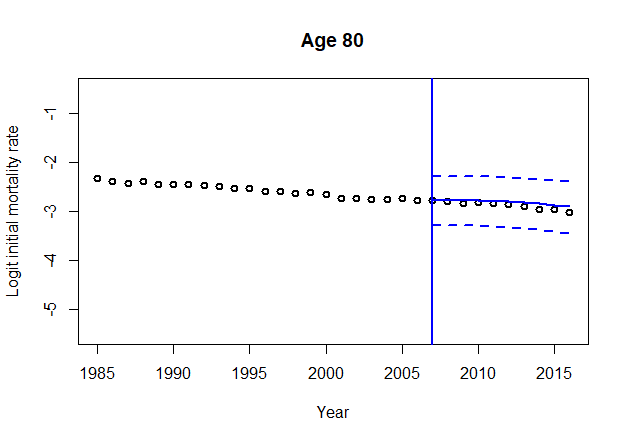}
	\end{minipage} 
	\begin{minipage}{0.5\textwidth}
		\centering
		\includegraphics[width=0.95\linewidth]{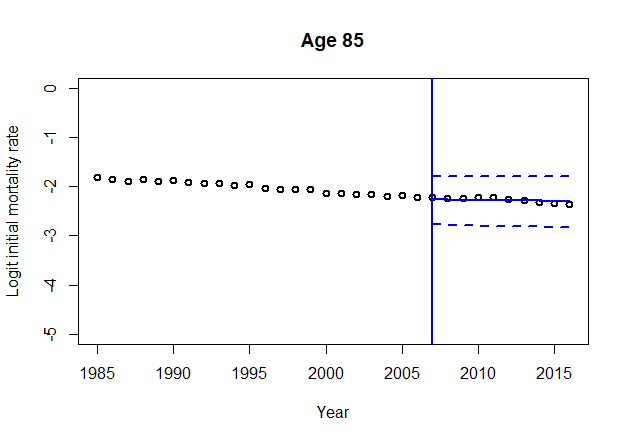}
	\end{minipage}
	\caption{Predicted logit initial male mortality rates of the selected age groups from ages 60 to 89 with 5-year age intervals using  \textcolor{black}{the proposed model} from the year 2007 to the year 2016 based on the observations from the year 1947 to the year 2006 in Japan. The circles are the observed values, the solid lines are the predictions, and the dashed lines are the 95\% prediction intervals. The vertical line indicates the start point of the predictions.} 
	\label{fig: Predicted logit total mortality odds of selected age groups from age 60 to age 89 with 5-year age intervals}
\end{figure}

\begin{figure}[!thb]
	\centering
	\includegraphics[width=0.95\linewidth]{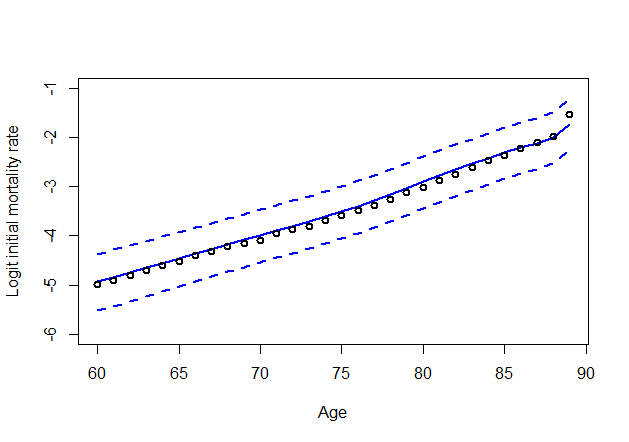}
	\caption{Predicted logit initial male mortality curve (with RMSE = 0.0793) from age 60 to age 89 with the 95\% prediction intervals using \textcolor{black}{the proposed model} for the year 2016 based on the observations from the year 1947 to the year 2006 in Japan. The circles are the observed logit initial male mortality rates, the solid line is the prediction and the dashed lines are the 95\% prediction intervals. }
	\label{fig: Predicted total mortality odds curve}
\end{figure}

 We fit the male mortality \textcolor{black}{data} in the training dataset into \textcolor{black}{the proposed model} stated in Equation (\ref{eq: (matrix) age group specific deviation trend function}). The mean components and random components of the proposed model can be estimated as discussed in Section \ref{Estimation of the time-series variant mixed-effects CBD model with age groups dependency and random cohort effects structure}. Thanks to the age group dependency design in the random components of the proposed model, we can \textcolor{black}{observe how} the mortality rates in different age groups correlate with each other from the estimated random effects intercept coefficients and \textcolor{black}{the} estimated random effects slope coefficients stated in Equation (\ref{eq: conditional mean of estimated gamma_1}) and Equation (\ref{eq: conditional mean of estimated gamma_2}) with the 95\% confidence intervals in Figure \ref{fig: Male estimated random effects intercept coefficients} and \ref{fig: Male estimated random effects slope coefficients}. We can see that they are both increasing along with the increase in age, indicating that age is a key factor to increase the male mortality rate. Figure \ref{fig: Male estimated random cohort effects coefficients} presents the estimated random cohort effects coefficients stated in Equation (\ref{eq: conditional mean of estimated gamma_3}). The shape of the estimated random cohort effects coefficients reflects the improvements or deteriorations of the male mortality among groups of people born within the same year. It shows that there is a significant improvement in mortality for those groups of people born in between the years 1880 to 1900, while a downturn happened for those groups of people born in around the years 1910 to 1920 in Japan. \textcolor{black}{This may be due to the fact that this group of people were sent to the main army force in the Second World War}. We then extrapolate the estimated random cohort effects coefficients for a 10-years-ahead out-of-sample forecast using (\ref{eq: extrapolation of the gamma_3}) \textcolor{black}{as shown in Figure \ref{fig: Male estimated random cohort effects coefficients}}. Figure \ref{fig: Predicted logit total mortality odds of selected age groups from age 60 to age 89 with 5-year age intervals} presents the predicted logit initial mate mortality rates of some selected age groups by the proposed method. We can see that the proposed model can capture the linear patterns of the logit initial mate mortality rates accurately with no significant misprediction. Figure \ref{fig: Predicted total mortality odds curve} gives an example of the 10-years-ahead forecast results of the logit initial mate mortality curves from age 60 to age 89 (with RMSE = 0.0590) with 95\% prediction intervals by the proposed approach for the year 2016 based on the observations from 1947 to 2006 in Japan. The root mean square error (RMSE) here is defined as
\[
\text{RMSE} = \sqrt{ \frac{1}{30} \sum_{j=1}^{30} \bigg( y_{x_{j},t_{2016}} - \hat{y}_{x_{j},t_{2016}} \bigg)^{2}},
\]  
where $y_{x_{j},t_{2016}}$ is the logit initial male mortality rate aged $x_{j}$ in the year 2016.
\subsection{Female mortality data} 
We move to the second case study of modelling and forecasting female mortality data. Female mortality data of Japan are also available from the year 1947 to the year 2016 in the \cite{human2017university}. We maintain the same selection for the age groups and also convert the central female mortality rate to the initial female mortality rate by Equation (\ref{eq: the relationship among q and m}) as in the male mortality case for demonstration in this section. 
\par
\begin{figure}[!thb]
	\centering
	\begin{minipage}{0.78\textwidth}
		\centering
		\includegraphics[width=1\linewidth]{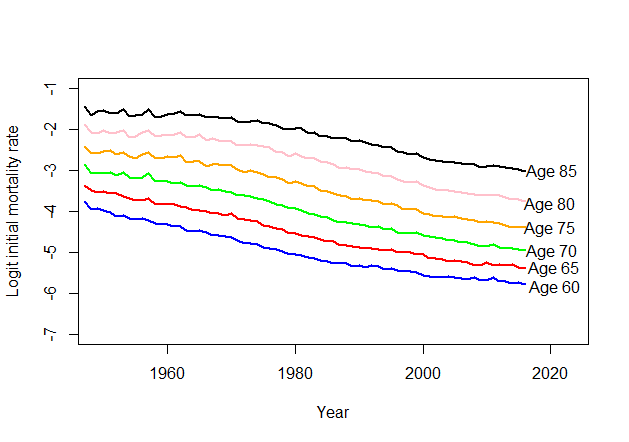}
		\subcaption{}
		\label{fig: Univariate time series of the observed logit initial female mortality rates with 5-year age intervals}
	\end{minipage}
	\begin{minipage}{0.78\textwidth}
		\centering
		\includegraphics[width=1\linewidth]{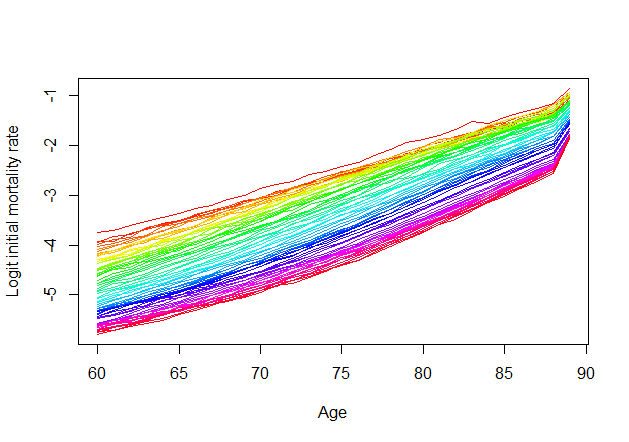}
		\subcaption{}
		\label{fig: Logit female mortality curves}
	\end{minipage}
	\caption{(a) Univariate time series of the observed logit initial female mortality rates with 5-year age intervals and (b) the logit initial female mortality curves from age 60 to age 89 from the year 1947 to the year 2016 in Japan.}
\end{figure}

 We present the historical \textcolor{black}{mortality} data of Japan as separate univariate time series of the logit initial female mortality rates with 5-year age intervals in Figure \ref{fig: Univariate time series of the observed logit initial female mortality rates with 5-year age intervals} and as the logit initial mortality curves from age 60 to age 89 from the year 1947 to the year 2016 in a time-ordering indicated by the colours of the rainbow from red to violet in Figure \ref{fig: Logit female mortality curves}. We can see that the downward movement of the time-series female mortality looks fairly similar among all the selected groups as well as the patterns of the male mortality over the examined period in Figure \ref{fig: Univariate time series of the observed logit initial female mortality rates with 5-year age intervals}. Figure \ref{fig: Logit female mortality curves} presents the same set of data from the age side as a bunch of the mortality curves in a time-ordering indicated by the colours of the rainbow from red to purple. Its vertical pattern shows the general trends and variations of the observed initial \textcolor{black}{female} mortality rates across the examined period. We can also see that the historical female mortality curves move downwards over time in general due mainly to the advances in medical technology and longer longevity as in the case of male mortality discussed in the previous section.
\subsection{Female mortality modelling and forecasting}
For the task of female mortality modelling and forecasting, we again attempt to make 10-years-ahead out-of-sample forecasts of the logit initial female mortality rates for demonstration and maintain the same setting for the split of dataset as for the male mortality case in this section.
\par
\begin{figure}[!thb]
	\begin{subfigure}{.5\textwidth}
		\centering
		\includegraphics[width=1\linewidth]{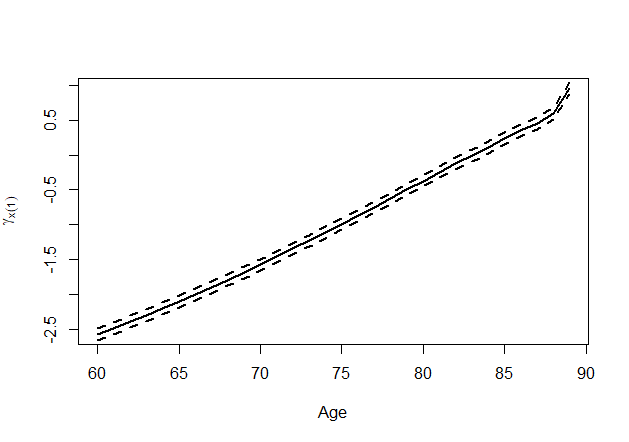}
		\subcaption{}
		\label{fig: female estimated random effects intercept coefficients}
	\end{subfigure}
	\begin{subfigure}{.5\textwidth}
		\centering
		\includegraphics[width=1\linewidth]{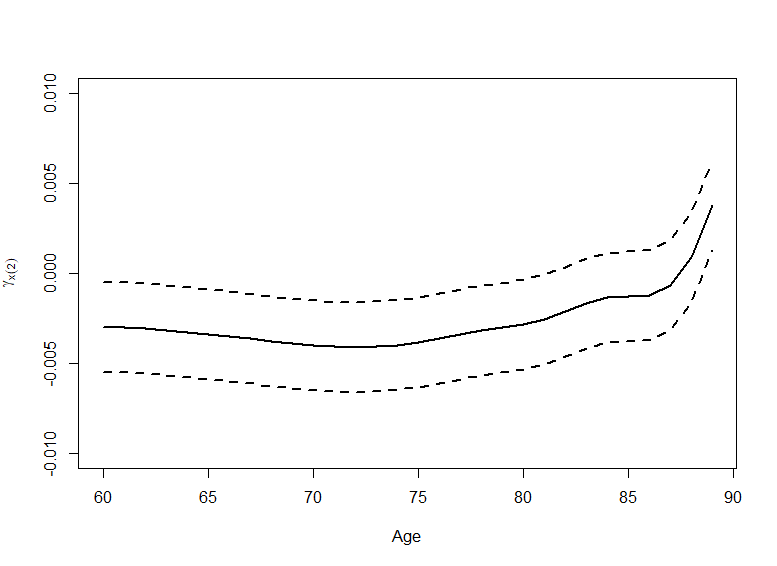}
		\subcaption{}
		\label{fig: female estimated random effects slope coefficients}
	\end{subfigure} \\[1ex]
	\begin{subfigure}{\linewidth}
		\centering
		\includegraphics[width=0.5\linewidth]{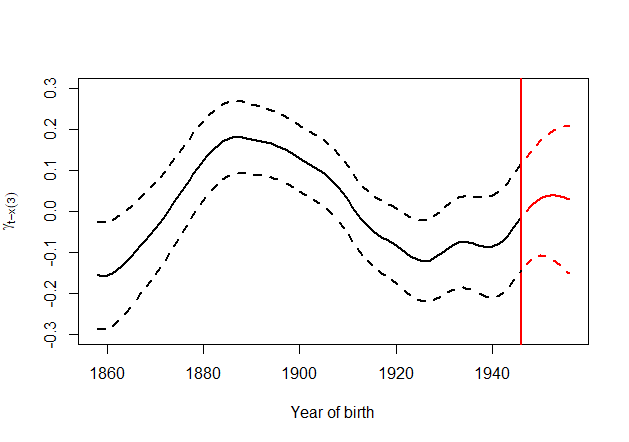}
		\subcaption{}
		\label{fig: female estimated random cohort effects coefficients}
	\end{subfigure}
	\caption{(a) Estimated random effects intercept coefficients, (b) the estimated random effects slope coefficients and (c) the estimated cohort effects coefficients using the observed logit initial female mortality rates of Japan from the year 1947 to the year 2006. Dashed lines are the 95\% confidence intervals and the vertical line indicates the start point of the 10-years-ahead forecasts of the estimated random cohort effects coefficients.}
	\label{fig: coefficients of the time-varaint mixed-effects CBD model}
\end{figure}

\begin{figure}[!thb]
	\begin{minipage}{0.5\textwidth}
		\centering
		\includegraphics[width=0.95\linewidth]{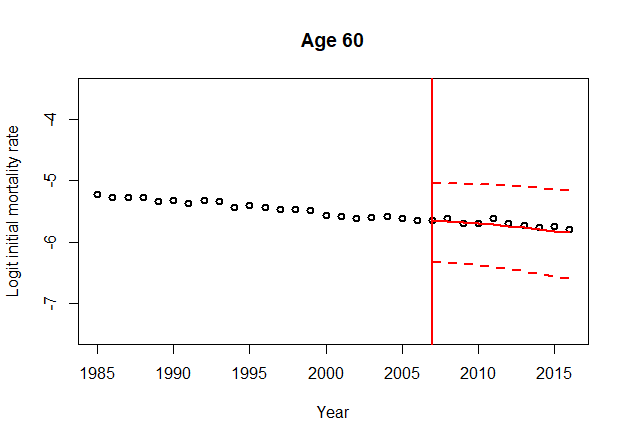}
	\end{minipage}
	\begin{minipage}{0.5\textwidth}
		\centering
		\includegraphics[width=0.95\linewidth]{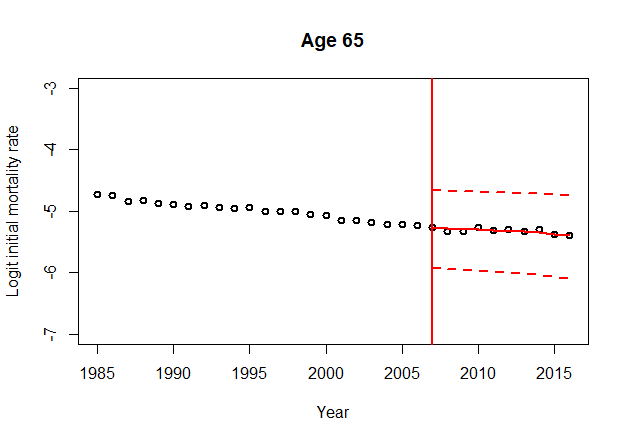}
	\end{minipage}
	\begin{minipage}{0.5\textwidth}
		\centering
		\includegraphics[width=0.95\linewidth]{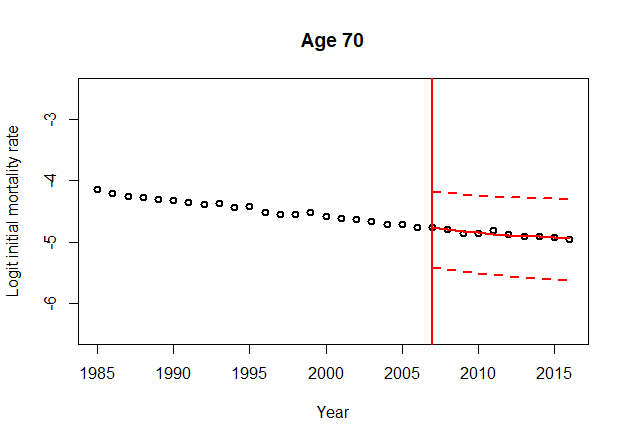}
	\end{minipage} 
	\begin{minipage}{0.5\textwidth}
		\centering
		\includegraphics[width=0.95\linewidth]{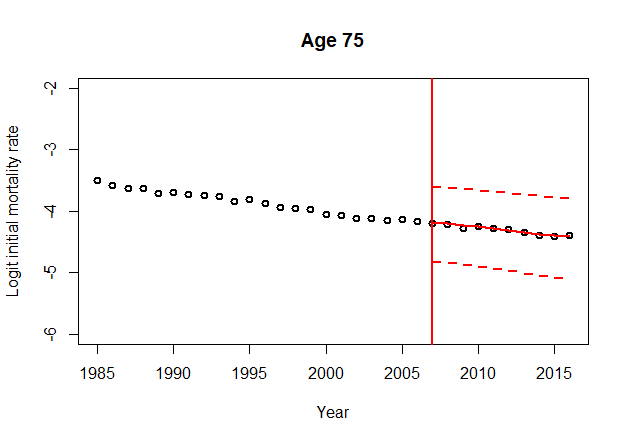}
	\end{minipage}
	\begin{minipage}{0.5\textwidth}
		\centering
		\includegraphics[width=0.95\linewidth]{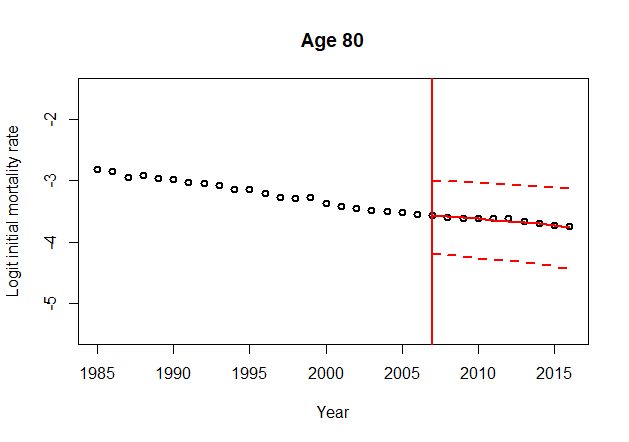}
	\end{minipage} 
	\begin{minipage}{0.5\textwidth}
		\centering
		\includegraphics[width=0.95\linewidth]{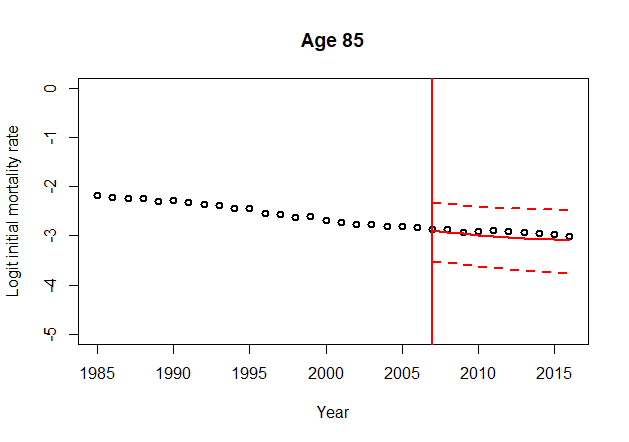}
	\end{minipage}
	\caption{Predicted logit initial male mortality rates of the selected age groups from age 60 to age 89 with 5-year age intervals using \textcolor{black}{the proposed model} from the year 2007 to the year 2016 based on the observations from the year 1947 to the year 2006 in Japan. The circles are the observed values, the solid lines are the predictions and the dashed lines are the 95\% prediction intervals. The vertical line indicates the start point of the predictions.} 
	\label{fig: Predicted logit initial female mortality rates of selected age groups from age 60 to age 89 with 5-year age intervals}
\end{figure}

\begin{figure}[!thb]
	\centering
	\includegraphics[width=0.95\linewidth]{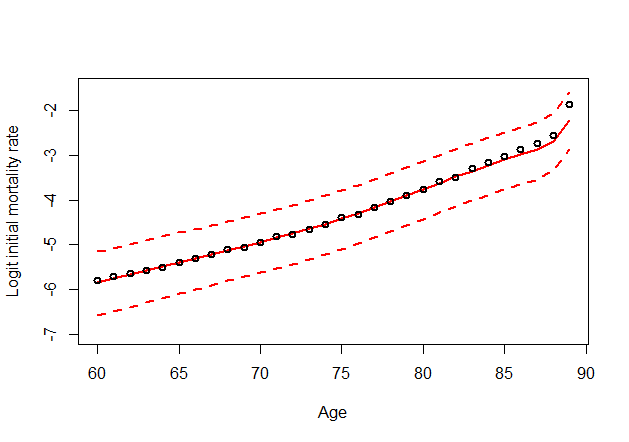}
	\caption{Predicted logit initial female mortality curve (with RMSE = 0.0842) from age 60 to age 89 with the 95\% prediction intervals using \textcolor{black}{the proposed model} for the year 2016 based on the observations from the year 1947 to the year 2006 in Japan. The circles are the observed logit initial female mortality rates, the solid line is the prediction and the dashed lines are the 95\% prediction intervals. }
	\label{fig: Predicted logit initial female mortality curve}
\end{figure}

 The female mortality  \textcolor{black}{data} in the training dataset is fitted into the proposed model stated in Equation (\ref{eq: (matrix) age group specific deviation trend function}). Figure \ref{fig: female estimated random effects intercept coefficients} and Figure \ref{fig: female estimated random effects slope coefficients} demonstrate the estimated random effects intercept coefficients and the estimated random effects slope coefficients with the 95\% confidence intervals for the fitted female mortality data. We can see that they also have increasing patterns along with the increase in age, reflecting that age is also an essential element to increase the female mortality rate in Japan. Figure \ref{fig: female estimated random cohort effects coefficients} presents the estimated random cohort effects coefficients stated in Equation (\ref{eq: conditional mean of estimated gamma_3}). The shape of the estimated random cohort effects coefficients reflects the improvements or deteriorations of the female mortality among groups of people born within the same year. Similar to the pattern of the estimated random cohort effects coefficients in the male mortality case, there is also a significant improvement in female mortality for those groups of people born in between the years 1880 to 1900, while a downturn happened for those groups of people born in around the years 1910 to 1920 in Japan. We then extrapolate the estimated random cohort effects coefficients for a 10-years-ahead out-of-sample forecast using Equation (\ref{eq: extrapolation of the gamma_3}) \textcolor{black}{as shown in Figure \ref{fig: female estimated random cohort effects coefficients}}. Figure \ref{fig: Predicted logit initial female mortality rates of selected age groups from age 60 to age 89 with 5-year age intervals} presents the predicted logit initial female mortality rates of the selected age groups by the proposed model. We can observe that the proposed model can catch  \textcolor{black}{the} linear patterns of the logit initial \textcolor{black}{female} mortality rates reasonably well with no significant mistake. Figure \ref{fig: Predicted logit initial female mortality curve} provides the demonstration of the 10-years-ahead forecast results of the logit initial  \textcolor{black}{female} mortality curves from age 60 to age 89 (with RMSE = 0.0842) with the 95\% prediction intervals by the proposed model for the year 2016 based on the observations from 1947 to 2006 in Japan.
\subsection{Forecast accuracy evaluations and comparisons with \textcolor{black}{the CBD model}}
We now compare and evaluate the forecast performance of \textcolor{black}{the proposed mixed-effects time series model} with \textcolor{black}{the CBD model} reviewed in Section \ref{Section 2: time-series variant mixed-effects CBD model}. In order to have a comprehensive investigation of the forecast accuracy among these two models, we consider ten other developed countries for which data are also available in the \cite{human2017university}. We restrict data periods of all selected countries commencing in the year 1947 up to the year 2016 (70 years in total) for a unified purpose. We examine and quantify the forecasting performance of the models by the rolling window analysis, which is frequently used for assessing the consistency of a \textcolor{black}{model's} forecasting ability by rolling a fixed size prediction interval (window) throughout the observed \textcolor{black}{period} \citep{zivot2007modeling}. We hold the sample data from the initial year up to the year $t_{l}$, where $t_{l} < t_{n}$, as holdout samples and produce the forecast for the $t_{l+h}$ year where $h$ is the forecast horizon. The forecasts errors are then determined by comparing the out-of-sample forecast results with the actual data. We increase one rolling-window (1 year ahead) in year $\textcolor{black}{t_{l+1}}$ to make the same procedure again for the year $t_{l+1+h}$ until the rolling-window analysis covers all available data in year $t_{n}$. We consider four different forecast horizons $(h = 5, 10, 15$ and $20)$ with ten sets of rolling-window to examine the short-term, the mid-term, and the long-term forecast abilities of the models. In our mortality rolling-window experiments, the \textcolor{black}{RMSE} is defined as 
\[
\text{RMSE}^{(i)}_{c}(h) = \sqrt{\frac{1}{10\times 30}\sum_{w=0}^{9}\sum_{j=1}^{30} \bigg( y^{(i)}_{x_{j}, t_{l+w+h}} - \hat{y}^{(i)}_{x_{j}, t_{l+w+h}} \bigg)^{2}},
\]   
where $c$ is the selected country, $w$ is the index of the rolling-window sets, $i$ is for male $(i = M)$ and for female $(i = F)$, and $j$ is the index of the examined age groups covering from age 60 to age 89.
\par Table \ref{table: Forecast accuracy of total mortality by the average RMSFEs in ten sets of rolling-windows analysis} presents the average RMSEs of ten sets of the rolling window analysis across the ten selected countries in four different forecast horizons in the male and female mortality curves fitting experiments. The back-testing results confirm that the forecasting performances and accuracy of \textcolor{black}{the proposed model} are much more desirable than \textcolor{black}{the CBD model}\footnote{\textcolor{black}{The CBD model} was implemented using \textit{R} package \textit{`StMoMo' in the numerical experiment} \citep{villegas2015stmomo}.}. Among all the ten selected countries and the four different forecast horizons, \textcolor{black}{the proposed model} scored the majority of the lowest RMSEs in both genders, different countries and forecast horizons in the numerical experiment.

\par 
\begin{table}[ht]
	\centering
	\begin{tabular}{ccc ccc cccc}  
		\toprule
		\multirow{1}{*}{
			\parbox[c]{.2\linewidth}{\centering Country}}
		& \multicolumn{3}{c}{\textcolor{black}{CBD model}} &&
		\multicolumn{3}{c}{\textcolor{black}{Proposed model}} \\
		\cmidrule{2-4} \cmidrule{6-8} \\
		& {\centering M} & {F} & {$\frac{\text{M+F}}{2}$} && {M} & {F} & {$\frac{\text{M+F}}{2}$} \\
		\midrule
		\underline{$h = 5$}\\
		Australia & \cellcolor{yellow}{\centering0.1174} & 0.1603  & {\centering 0.1388} &&
		0.1398 & \cellcolor{yellow}0.1103& \cellcolor{yellow}0.1251
\\
		Belgium & {\centering0.1039} & 0.1547  & {\centering 0.1293} &&
		\cellcolor{yellow}0.0783 & \cellcolor{yellow}0.0557& \cellcolor{yellow}0.0670
		 \\
		
		Canada & {\centering0.0938} & 0.1607  & {\centering 0.1272} &&
		\cellcolor{yellow}0.0481 & \cellcolor{yellow}0.0547& \cellcolor{yellow}0.0514\\
		
		France & {\centering0.1929} & {\centering0.2614} & 0.2271 &&
		\cellcolor{yellow}0.0780 & \cellcolor{yellow}0.0580 & \cellcolor{yellow}0.0680\\
		
		Italy & \cellcolor{yellow}{\centering0.1076} & 0.1544  & {\centering 0.1310} &&
		0.1109 & \cellcolor{yellow}0.0624 & \cellcolor{yellow}0.0866\\
		
		Japan & {\centering0.1229} & 0.1659  & {\centering 0.1444} &&
		\cellcolor{yellow}0.0682 & \cellcolor{yellow}0.0759 & \cellcolor{yellow}0.0721\\
		
		Netherlands & {\centering0.1219} & 0.1769  & {\centering 0.1494} &&
		\cellcolor{yellow}0.1050 & \cellcolor{yellow}0.1423 & \cellcolor{yellow}0.1236\\
		
		Spain & {\centering0.1106} & 0.1570  & {\centering 0.1338} &&
		\cellcolor{yellow}0.0839 & \cellcolor{yellow}0.0744 & \cellcolor{yellow}0.0792\\ 
		
		UK & {\centering0.1381} & 0.1280  & {\centering 0.1330}  &&
		\cellcolor{yellow}0.1237 & \cellcolor{yellow}0.0584 & \cellcolor{yellow}0.0911\\ 
		
		USA & {\centering0.1065} & 0.1729  & {\centering 0.1397} &&
		\cellcolor{yellow}0.0558 & \cellcolor{yellow}0.0600 & \cellcolor{yellow}0.0579\\
		
		Average & {\centering0.1215} & 0.1692  & {\centering 0.1454} &&
		\cellcolor{yellow}\textbf{0.0892} & \cellcolor{yellow}\textbf{0.0752} & \cellcolor{yellow}\textbf{0.0822}\\
		\midrule
		\underline{$h = 10$}\\
		
		Australia &  \cellcolor{yellow}{\centering0.1644} & 0.2270 & 0.1957  && 0.1660 & \cellcolor{yellow}0.1635 & \cellcolor{yellow}0.1648  \\
		
		Belgium &  0.1196 & 0.2054 & 0.1625  && \cellcolor{yellow}0.1153
		 & \cellcolor{yellow}0.1467 & \cellcolor{yellow}0.1310\\
		
		Canada &  0.1337 & 0.2467 & 0.1902 && \cellcolor{yellow}0.1042
		& \cellcolor{yellow}0.1174 & \cellcolor{yellow}0.1108   \\
		
	    France &  0.3247 & 0.4384 & 0.3815 && \cellcolor{yellow}0.0787
	    & \cellcolor{yellow}0.0846 & \cellcolor{yellow}0.0817   \\
	    
		Italy &  0.1562 & 0.2079 & 0.1821 && \cellcolor{yellow}0.1194
		& \cellcolor{yellow}0.0731 & \cellcolor{yellow}0.0963   \\
		
		Japan &  0.1756 & 0.2093 & 0.1925  && \cellcolor{yellow}0.1023
		& \cellcolor{yellow}0.1134 & \cellcolor{yellow}0.1078   \\
		
        Netherlands &  0.1580 & 0.2572 & 0.2076 && \cellcolor{yellow}0.1504 & \cellcolor{yellow}0.1765 & \cellcolor{yellow}0.1635  \\
        
		Spain &  0.1563 & 0.2234 & 0.1899 && \cellcolor{yellow}0.1191
		& \cellcolor{yellow}0.1039 & \cellcolor{yellow}0.1115  \\
	
		UK &  0.1915 & 0.1812 & 0.1864 && \cellcolor{yellow}0.1360
		& \cellcolor{yellow}0.1493 & \cellcolor{yellow}0.1426 \\
		
		USA & 0.1735 & 0.2953 & 0.2344 && \cellcolor{yellow}0.1308
		& \cellcolor{yellow}0.0742 & \cellcolor{yellow}0.1025 \\
		
	    Average &  0.1754 & 0.2492 & 0.2123 && \cellcolor{yellow}\textbf{0.1222}
	    & \cellcolor{yellow}\textbf{0.1203} & \cellcolor{yellow}\textbf{0.1213} \\
		\bottomrule
	\end{tabular}
	\caption{The average RMSEs in ten sets of the rolling-windows analysis of \textcolor{black}{the predicted} mortality curves across the ten selected countries using \textcolor{black}{the CBD model} and \textcolor{black}{the proposed model}. The minimal forecast errors are highlighted, and the lowest averaged forecast errors among models in different forecast horizons are highlighted in bold.} 
	\label{table: Forecast accuracy of total mortality by the average RMSFEs in ten sets of rolling-windows analysis}
\end{table}

\par 
\begin{table}[ht]
	\ContinuedFloat
	\centering
	\begin{tabular}{ccc ccc cccc}  
	\toprule
	\multirow{1}{*}{
		\parbox[c]{.2\linewidth}{\centering Country}}
	& \multicolumn{3}{c}{\textcolor{black}{CBD model}} &&
	\multicolumn{3}{c}{\textcolor{black}{Proposed model}} \\
	\cmidrule{2-4} \cmidrule{6-8} \\
	& {\centering M} & {F} & {$\frac{\text{M+F}}{2}$} && {M} & {F} & {$\frac{\text{M+F}}{2}$} \\
\midrule
\underline{$h = 15$}\\
Australia &  \cellcolor{yellow}0.2373 & 0.2679 & 0.2526 && 
0.2899 & \cellcolor{yellow}0.2039 & \cellcolor{yellow}0.2469\\

Belgium &  0.1412 & 0.2126 & 0.1769 && 
\cellcolor{yellow}0.1373 & \cellcolor{yellow}0.1286 & \cellcolor{yellow}0.1330\\

Canada &  0.1753 & 0.2851 & 0.2302 && 
\cellcolor{yellow}0.1373 & \cellcolor{yellow}0.0631 & \cellcolor{yellow}0.1002\\

France &  0.3858 & 0.4793 & 0.4326 && 
\cellcolor{yellow}0.1672 & \cellcolor{yellow}0.1009 & \cellcolor{yellow}0.1340\\

Italy &  \cellcolor{yellow}0.2212 & 0.2460 & 0.2336  && 
0.2446 & \cellcolor{yellow}0.0899 & \cellcolor{yellow}0.1672\\

Japan &  0.2271 & 0.2268 & 0.2270  && 
\cellcolor{yellow}0.1504 & \cellcolor{yellow}0.1079 & \cellcolor{yellow}0.1292\\

Netherlands &  0.2454 & 0.2680 & 0.2567 && 
\cellcolor{yellow}0.2276 & \cellcolor{yellow}0.1969 & \cellcolor{yellow}0.2123\\

Spain &  0.2092 & 0.2848 & 0.2470 && \cellcolor{yellow}0.1536
& \cellcolor{yellow}0.1360 &  \cellcolor{yellow}0.1448    \\

UK &  0.2698 & 0.2300 & 0.2499  && \cellcolor{yellow}0.1540
& \cellcolor{yellow}0.2130 &  \cellcolor{yellow}0.1835    \\

USA & 0.2056 & 0.3218 & 0.2637 && \cellcolor{yellow}0.1153
& \cellcolor{yellow}0.0523 & \cellcolor{yellow}0.0838    \\

Average &  0.2318 & 0.2822 & 0.2570 && \cellcolor{yellow} \textbf{0.1777}
& \cellcolor{yellow}\textbf{0.1292} & \cellcolor{yellow}\textbf{0.1535}  \\
\midrule
\underline{$h = 20$}\\
Australia &  \cellcolor{yellow}0.3387 & 0.2887 & \cellcolor{yellow}0.3137  && 
0.3878 & \cellcolor{yellow}0.2783 & 0.3330\\

Belgium &  \cellcolor{yellow}0.1835 & 0.2280 & 0.2057 && 
0.2052 & \cellcolor{yellow}0.1462 & \cellcolor{yellow}0.1757\\

Canada &  \cellcolor{yellow}0.2250 & 0.2680 & 0.2465  && 
0.3091 & \cellcolor{yellow}0.0582 & \cellcolor{yellow}0.1836\\

France &  0.4166 & 0.4128 & 0.4147   && 
\cellcolor{yellow}0.1876 & \cellcolor{yellow}0.1287 & \cellcolor{yellow}0.1582\\

Italy &  0.3180 & 0.3037 & 0.3109    && 
\cellcolor{yellow}0.2875 & \cellcolor{yellow}0.1204 & \cellcolor{yellow}0.2039\\

Japan &  0.2945 & 0.2127 & 0.2536    && 
\cellcolor{yellow}0.2551 & \cellcolor{yellow}0.1258 & \cellcolor{yellow}0.1904\\

Netherlands &  0.3457 & 0.2922 & 0.3190 && 
\cellcolor{yellow}0.3111 & \cellcolor{yellow}0.1729 & \cellcolor{yellow}0.2420\\

Spain &  0.2628 & 0.3555 & 0.3091      && 
\cellcolor{yellow}0.1972 & \cellcolor{yellow}0.1787 & \cellcolor{yellow}0.1880\\

UK &  0.3711 & 0.2950 & 0.3331  && 
\cellcolor{yellow}0.2866 & \cellcolor{yellow}0.2538 & \cellcolor{yellow}0.2702\\

USA & 0.2173 & 0.2906 & 0.2539   && 
\cellcolor{yellow}0.1706 & \cellcolor{yellow}0.0532 & \cellcolor{yellow}0.1119\\

Average &  0.2973 & 0.2947 & 0.2960   && 
\cellcolor{yellow}\textbf{0.2598} & \cellcolor{yellow}\textbf{0.1516} & \cellcolor{yellow}\textbf{0.2057}\\ 
\bottomrule
\end{tabular}
	\caption{ \textit{(continued)} The average RMSEs in ten sets of the rolling-windows analysis of the \textcolor{black}{predicted} mortality curves across the ten selected countries using \textcolor{black}{the CBD model} and \textcolor{black}{the proposed model}. The minimal forecast errors are highlighted, and the lowest averaged forecast errors among models in different forecast horizons are highlighted in bold.} 
\end{table}
\section{Discussion and conclusion remarks} \label{Section 5: time-series variant mixed-effects CBD model }
\par In this article, we have introduced \textcolor{black}{a novel mortality model} under the time-series and mixed-effects \textcolor{black}{structure}. We have presented the formulation, the model assumptions and the estimation of the parameters \textcolor{black}{as well as the differences} between \textcolor{black}{the CBD model} and \textcolor{black}{the proposed model}. We lastly compared the forecasting abilities among the two models using the rolling window analysis across the mortality data of ten selected countries. \textcolor{black}{The proposed mixed effects model} has four remarkable advantages over \textcolor{black}{the CBD} model \textcolor{black}{as} summarised \textcolor{black}{below.}
\begin{enumerate}
\item[\textbf{1.}]{\textbf{Circumventing the age-period-cohort identification problem}}\\
\textcolor{black}{The proposed mixed effects model} provides an alternative way to avoid the age-period-cohort identification problem when estimating the model cohort effects parameters. Unlike \textcolor{black}{the CBD model which estimates} the model cohort effects parameters as the fixed-effects parameters by placing the specified constraints through the iterative optimisation process, \textcolor{black}{the proposed model} under the mixed-effects structure can produce a unique best fitting solution without the need for introducing any pre-defined constraints on the model cohort effects parameters. It is because the model cohort effects parameters are treated as random variables with their own specific distributions. The structure of \textcolor{black}{the proposed model} also allows us to examine how different age groups correlate with each other and disclose underlying mortality information about the age groups correlation patterns in a more compendious fashion than \textcolor{black}{the CBD model}.
	
\item[\textbf{2.}]{\textbf{Unified time series framework for out-of-sample extrapolation}}\\
\textcolor{black}{The CBD model} \textcolor{black}{obtains} the estimates of the model parameters under a fixed-effects regression based approach first, then \textcolor{black}{applies} the random walk with drift processes method to do the model coefficients extrapolation in the second step. \textcolor{black}{In contract, our proposed model} can extrapolate the out-of-sample mortality rates under a unified single step framework. Indeed, \cite{de2006extending}, \cite{sweeting2011trend}, \cite{fung2017unified} and \textcolor{black}{\cite{li2017semi}} doubt about the appropriateness of using the random walk \textcolor{black}{with drift processes} method to extrapolate the coefficients of the CBD model as a two-step approach. \cite{leng2016inference} also argue that the two-steps approach is a somewhat ad-hoc procedure, \textcolor{black}{and} it would be more ideal to perform the estimation and the extrapolation under a consistent and rigorous single universal framework from the statistical point of view. \textcolor{black}{The proposed model} in this article aims to provide \textcolor{black}{the unified time series framework} and avoids any potential pitfalls using the two-step estimation procedure that \textcolor{black}{the CBD model} may fall.
	
\item[\textbf{3.}]{\textbf{ \textcolor{black}{Natural} confidence intervals under the normal distribution assumption}}\\
\textcolor{black}{The CBD model} is constructed based on the Poisson distribution assumption,  \textcolor{black}{and it is difficult to quantify} the uncertainties of \textcolor{black}{the} estimated model  \textcolor{black}{parameters due} to the equal Poisson mean and variance and the over-dispersion problems. \textcolor{black}{However}, the \textcolor{black}{proposed model} can provide natural confidence intervals that include the estimated model parameters uncertainties and the prediction intervals for the out-of-sample extrapolation given that the proposed model is established under the Bayesian paradigm with the normal distribution assumption. The ability to quantifying the parameter risks can enhance the reliability of the prediction, and this ability is especially practical for valuing the uncertainties \textcolor{black}{for} life insurance portfolios and pension schemes in the insurance industry.
	
\item[\textbf{4.}]{\textbf{\textcolor{black}{Improved} forecasting accuracy}}\\
The forecasting performance of \textcolor{black}{the proposed model} is better than \textcolor{black}{the CBD model}. In our window-rolling \textcolor{black}{experiments} across the mortality data of \textcolor{black}{the ten} developed countries, the accuracies of the \textcolor{black}{predicted} mortality curves are significantly improved by \textcolor{black}{the proposed model} in short-term, mid-term and long-term forecasting. The improvements may be mainly due to the time-series and mixed-effects framework of \textcolor{black}{the proposed model}, which can directly extrapolate the data forward without the need of placing any extra assumptions on the model parameters.
\end{enumerate}
\par Despite the advantages of \textcolor{black}{the proposed CBD model} listed above, \textcolor{black}{both our model and the CBD model} only work when there is a linear pattern in the latter part of the mortality curve above the pension age. \textcolor{black}{Neither of them are} capable of handling any non-linear \textcolor{black}{mortality} patterns for young \textcolor{black}{ages}. Although some attempts have been made to extend the \textcolor{black}{existing} CBD model design while keeping its simplicity to cover the whole age range for mortality modelling, see, for example, \cite{plat2009stochastic}, \cite{cairns2009quantitative}, \textcolor{black}{further} research is still needed to consider about extending the proposed model with more factors which can handle more comprehensive age ranges or embedding some time weightings approaches in the model estimation procedure so that it can capture more recent mortality patterns for better forecasting performance.
\begin{spacing}{1.5}
	\bibliographystyle{apacite}
	\bibliography{reference}
\end{spacing}
\end{document}